\documentclass[a4paper,fleqn,usenatbib]{mnras}
\usepackage[T1]{fontenc}
\usepackage{ae,aecompl}
\usepackage{graphicx}
\usepackage{amsmath}
\usepackage{amssymb}
\usepackage{txfonts}
\usepackage{slashed}
\usepackage{makecell}
\usepackage{capt-of}
\usepackage{bm}
\usepackage{subcaption}
\usepackage{multicol}
\usepackage{caption}

\usepackage[dvipsnames]{xcolor}
\definecolor{armygreen}{rgb}{0.0, 0.6, 0.4}

\hypersetup{pdfauthor={A. Spurio Mancini \textit{et el.}},
               pdftitle={KiDS:GAMA: constraints on Horndeski gravity},
              pdfkeywords={gravitational lensing: weak -- dark energy},
               bookmarksnumbered=true}

\usepackage{etoolbox}
\makeatletter
\patchcmd\@combinedblfloats{\box\@outputbox}{\unvbox\@outputbox}{}{
  \errmessage{\noexpand\@combinedblfloats could not be patched}
}
\makeatother

\newcommand{\ltsima}{$\; \buildrel < \over \sim \;$}
\newcommand{\lsim}{\lower.5ex\hbox{\ltsima}}
\newcommand{\gtsima}{$\; \buildrel > \over \sim \;$}
\newcommand{\gsim}{\lower.5ex\hbox{\gtsima}}

\usepackage[section]{placeins}
\usepackage{cleveref}

\title[KiDS+GAMA: constraints on Horndeski gravity]{KiDS+GAMA: Constraints on Horndeski gravity from combined large-scale structure probes}

\author[A. Spurio Mancini et al.]{
A. Spurio Mancini,$^{1,2,3,4}$\thanks{Contact e-mail: \href{mailto:as2848@cam.ac.uk}{as2848@cam.ac.uk}}
F. K\"{o}hlinger,$^{5}$
B. Joachimi,$^{4}$
V. Pettorino,$^{6}$
B. M. Sch\"{a}fer,$^{7}$
\newauthor 
R. Reischke,$^{8, 7}$
E. van Uitert,$^{4}$
S. Brieden,$^{9,10}$
M. Archidiacono,$^{11}$
J. Lesgourgues$^{11}$
\\
$^{1}$Kavli Institute for Cosmology, Cambridge, Madingley Road, Cambridge, CB3 0HA, UK\\
$^{2}$Astrophysics Group, Cavendish Laboratory, J. J. Thomson Avenue, Cambridge, CB3 0HE, UK\\
$^{3}$Institut f{\"u}r Theoretische Physik, Universit{\"a}t Heidelberg, Philosophenweg 12, 69120 Heidelberg, Germany\\
$^{4}$Department of Physics and Astronomy, University College London, Gower Street, London WC1E 6BT, UK\\
$^{5}$Kavli Institute for the Physics and Mathematics of the Universe (Kavli IPMU, WPI), The University of Tokyo \\
Institutes for Advanced Study, The University of Tokyo, Kashiwa, Chiba 277-8583, Japan\\
$^{6}$AIM, CEA, CNRS, Universit\'{e} Paris-Saclay, Universit\'{e} Paris Diderot, Sorbonne Paris Cit\'{e}, F-91191 Gif-sur-Yvette, France\\
$^{7}$Astronomisches Rechen-Institut, Zentrum f{\"u}r Astronomie der Universit{\"a}t Heidelberg, Philosophenweg 12, 69120 Heidelberg, Germany\\
$^{8}$Physics Department, Technion University, Haifa 32000, Israel\\
$^{9}$ICC, University of Barcelona, IEEC-UB, Mart\'{i} i Franqu\'{e}s, 1, 08028 Barcelona, Spain\\
$^{10}$Dept. de F\'{i}sica Qu\`{a}ntica i Astrof\'{i}sica, Universitat de Barcelona, Mart\'{i} i Franqu\'{e}s, 1, 08028 Barcelona, Spain\\
$^{11}$Institute for Theoretical Particle Physics and Cosmology (TTK), RWTH Aachen University, D-52056 Aachen, Germany}
\date{}

\begin{document}
\pagerange{\pageref{firstpage}--\pageref{lastpage}}
\pubyear{2018}
\maketitle
\label{firstpage}

\setlength{\parindent}{10pt}

\begin{abstract}
We present constraints on Horndeski gravity from a combined analysis of cosmic shear, galaxy-galaxy lensing and galaxy clustering from $450\,\mathrm{deg}^2$ of the Kilo-Degree Survey (KiDS) and the Galaxy And Mass Assembly (GAMA) survey. The Horndeski class of dark energy/modified gravity models includes the majority of universally coupled extensions to $\Lambda$CDM with one scalar field in addition to the metric. We study the functions of time that fully describe the evolution of linear perturbations in Horndeski gravity. Our results are compatible throughout with a $\Lambda$CDM model. By imposing gravitational wave constraints, we fix the tensor speed excess to zero and consider a subset of models including e.g. quintessence and $f(R)$ theories. Assuming proportionality of the Horndeski functions $\alpha_B$ and $\alpha_M$ (kinetic braiding and the Planck mass run rate, respectively) to the dark energy density fraction $\Omega_{\mathrm{DE}}(a) = 1 - \Omega_{\mathrm{m}}(a)$, we find for the proportionality coefficients $\hat{\alpha}_B = 0.20_{-0.33}^{+0.20} \,$ and $\, \hat{\alpha}_M = 0.25_{-0.29}^{+0.19}$. Our value of $S_8 \equiv \sigma_8 \sqrt{\Omega_{\mathrm{m}}/0.3}$ is in better agreement with the Planck estimate when measured in the enlarged Horndeski parameter space than in a pure $\Lambda$CDM scenario. In our joint three-probe analysis we report a downward shift of the $S_8$ best fit value from the \textit{Planck} measurement of $\Delta S_8 = 0.016_{-0.046}^{+0.048}$ in Horndeski gravity, compared to $\Delta S_8 = 0.059_{-0.039}^{+0.040}$ in $\Lambda$CDM. Our constraints are robust to the modelling uncertainty of the non-linear matter power spectrum in Horndeski gravity. Our likelihood code for multi-probe analysis in both $\Lambda$CDM and Horndeski gravity is publicly available at \url{https://github.com/alessiospuriomancini/KiDSHorndeski}.
\end{abstract}


\section{Introduction}

The investigation of the accelerated expansion of the Universe is one of the main areas of active research in cosmology. A cosmological constant $\Lambda$ is an excellent fit to most observations and can be considered responsible for cosmic acceleration. However, while representing a key component of the concordance $\Lambda$CDM model, the cosmological constant still lacks of a solid theoretical understanding in terms of naturalness and interpretation as the energy density of the vacuum \citep[see e.g.][for a review]{Martin2012}. As an alternative to the cosmological constant, a dark energy component has been invoked, as a fluid with sufficiently negative pressure to drive the accelerated expansion of the Universe. Another possible interpretation of the observed acceleration may be as the consequence of a modification of the laws of gravity on large, cosmological scales, marking deviations from General Relativity \citep[see e.g.][for a review]{Clifton2012}. 

A large fraction of models for dark energy and modified gravity is characterized by the presence of a scalar field in the Lagrange density, as an additional gravitational degree of freedom parallel to the metric. The mathematical formulation of the most general expression for the Lagrange density of such a scalar-tensor theory, with derivatives in the equations of motion not higher than second order, was first discussed in \citet{Horndeski1974} and subsequently rediscovered in \citet{Nicolis2009} and \citet{Deffayet2011}. This so-called Horndeski Lagrangian encompasses a remarkably large number of dark energy and modified gravity models (see Section~\ref{sec:Horndeski} for a list of some of them); the condition on the highest order of the derivatives guarantees the stability of all these theories against ghost-like degrees of freedom \citep[Ostrogradsky's instabilities; see][]{Woodard2015}.

Current and future surveys aim at using different cosmological probes to investigate the true nature of the accelerated expansion \citep[see][for a review]{Weinberg2013}, with weak gravitational lensing and galaxy clustering at the forefront of these studies. 

The weak gravitational lensing effect is particularly sensitive to the growth of cosmic structure, encoding therefore precious information on the acceleration as a dynamical effect in redshift, or equivalently time. Particularly rich in information on dark energy and modified gravity is the weak lensing effect produced by the large-scale structure (LSS) of the Universe, or \textit{cosmic shear}. This is observed in the form of correlations of galaxy shapes due to the distortion of the cross-sectional shape of light bundles coming from background sources, caused by the gravitational tidal fields of the LSS in the foreground \citep[see e.g.][for reviews on the topic]{Bartelmann2001, Hoekstra2008, Kilbinger2015}. 

Besides considering correlations of galaxy shapes, information on the weak lensing effect can be extracted correlating the positions of foreground galaxies, which trace the large-scale structure, with the shapes of background galaxies. This galaxy-matter cross-correlation is often referred to as \textit{galaxy-galaxy lensing}.

The shape distortions produced by weak gravitational lensing are difficult to measure, since the induced source galaxy ellipticities are at the percent level, and a number of systematic effects can mimic this signal \citep{Mandelbaum2017}. Despite these difficulties, cosmological results have already been derived by numerous cosmic shear surveys \citep{Heymans2013, Jee2013, Hildebrandt2017, Hikage2018, Troxel2018}. Interestingly, the fiducial cosmic shear analyses of the Canada-France-Hawaii Lensing Survey \footnote{\url{http://www.cfhtlens.org/}}\citep[CFHTLenS;][]{Heymans2013, Joudaki2017} and the Kilo-Degree Survey \footnote{\url{http://kids.strw.leidenuniv.nl/}}\citep[KiDS;][]{Hildebrandt2017, Kohlinger2017}, both prefer a cosmological model that is in mild tension with the best-fitting parameters obtained from Cosmic Microwave Background (CMB) measurements of the \textit{Planck} \footnote{\url{http://sci.esa.int/planck/}} satellite \citep{Planck2016_XIII, PlanckCollaboration2016, Planck2018}. 

Galaxy-galaxy lensing measurements have also matured in recent years and can help constrain dark energy \citep[see e.g.][]{Kwan2017}, in particular when combined with studies of \textit{galaxy clustering} \citep[see e.g. the recent analysis of][]{Singh2018}: the statistic describing at lowest order the galaxy spatial distribution is the two-point correlation function, which in the past provided early evidence for the $\Lambda$CDM model \citep{Baugh1996, Eisenstein2001, Saunders2000, Huterer2001, Hamilton2002, Cole2005, Tegmark2006}.~Today, studying the spatial distribution of galaxies and its evolution in time is crucial to analyse possible extensions to the cosmological concordance model. The evolution of the clustering of galaxies with redshift can put direct constraints on models for the evolution of density perturbations, which is key to discriminate between different dark energy/modified gravity theories. However, the interpretation of galaxy clustering is complicated by galaxy bias \citep{Kaiser1984}, the relation between the galaxy spatial distribution and the theoretically predicted matter distribution.

Modern optical imaging surveys measure the positions and ellipticities of millions of galaxies; from them, the galaxy overdensity field as well as the gravitational lensing shear field can be derived. The two-point auto- and cross-correlations of these two fields are the two-point correlation functions of cosmic shear, galaxy-galaxy lensing and galaxy clustering. A joint analysis of these correlation functions can break degeneracies between cosmological and nuisance parameters, leading to tighter cosmological constraints \citep{Joachimi2010}. Several earlier studies have indeed considered such joint analyses \citep{Cacciato2013, Mandelbaum2013, More2015, Kwan2017, Nicola2017}, albeit very few in a modified gravity context. Among the latter, \citet{Joudaki2018} recently performed a combined analysis of cosmic shear tomography, galaxy-galaxy lensing tomography, and redshift-space multipole power spectra using data from KiDS-450 ($\sim 450~\mathrm{deg}^2$ of cosmic shear data from the KiDS survey) and two overlapping spectroscopic surveys, the 2-degree Field Lensing Survey \footnote{\url{http://2dflens.swin.edu.au/}} \citep[2dFLenS;][]{Blake2016} and the Baryon Oscillation Spectroscopic Survey \footnote{\url{http://www.sdss3.org/surveys/boss.php}} \citep[BOSS;][]{Dawson2013}. They found that none of the extended cosmologies considered were simultaneously favoured in a model selection sense and able to resolve the discordance with \textit{Planck}, except for an evolving dark energy component with a time-dependent $w_0 - w_a$ equation of state. \citet{Amon2018} presented a measurement of $E_G$, a statistic combining measurements of weak gravitational lensing, galaxy clustering and redshift space distortions, proposed as a consistency test of General Relativity \citep{Zhang2007}. They determined the value of $E_G$ using data from the KiDS, 2dFLenS, BOSS and Galaxy And Mass Assembly \footnote{\url{http://www.gama-survey.org/}} \citep[GAMA;][]{Driver2009, Driver2011, Liske2015} surveys; their results show that measurements of the $E_{\mathrm{G}}$ statistic cannot be conducted as consistency checks of General Relativity until the aforementioned tension in cosmological parameters is resolved, and their $E_{\mathrm{G}}$ measurements favour a lower matter density cosmology than the CMB. Recently, \citet{Abbott2018} presented a combined analysis of galaxy clustering and weak gravitational lensing from the first-year data of the Dark Energy Survey, targeting modifications of the metric potentials that would be a signal of modified gravity. They found that their constraints are compatible with a cosmological constant scenario.

For a flat $\Lambda$CDM model, \citet[][hereafter vU18]{vanUitert2018} exploited the overlap of the KiDS and GAMA surveys to produce constraints on cosmological parameters by cross-correlating cosmic shear measurements from KiDS-450, the galaxy-matter cross-correlation from KiDS-450 around two foreground samples of GAMA galaxies, and the angular correlation function of the same foreground galaxies. Their results are consistent with both the KiDS-450 \citep{Hildebrandt2017, Kohlinger2017} and \textit{Planck} analyses \citep{Planck2016_XIII, PlanckCollaboration2016, Planck2018} when considering the cross-correlation of all three probes; with cosmic shear alone, their results are fully consistent with the KiDS analysis, while showing a similar tension with the \textit{Planck} results. 

In this work we extend the analysis of vU18 to the Horndeski class of dark energy and modified gravity models. The evolution of linear cosmological perturbations in Horndeski gravity can be fully described by four functions of time only \citep{Gleyzes2013, BelliniSawicki2014}. Assuming a time parameterization for these functions, it is possible to set constraints on the parameters describing their time evolution: this is the main goal of our study. We consider two parameterizations and concentrate in particular on a parameterization that sets these functions proportional to the dark energy density fraction $\Omega_{\mathrm{DE}}(a) = 1 - \Omega_{\mathrm{m}}(a)$; this choice allows the Horndeski functions to reproduce the late-time cosmic acceleration. One of these functions is largely unconstrained by large-scale structure probes, and uncorrelated with all other cosmological parameters, therefore we fix its value. We fix also the Horndeski function describing deviations of the gravitational wave speed from that of light, as it has been recently constrained by gravitational wave detections. We set constraints on the remaining two Horndeski functions, describing how the scalar and
metric kinetic terms mix and how the effective gravitational constant evolves over time, respectively.

This paper is structured as follows. In Section~\ref{sec:Horndeski} we review the Horndeski class of dark energy/modified gravity models and introduce the description of linear perturbations underlying our constraints. In Section~\ref{sec:theory} we present expressions for the power spectra of the three probes considered, assuming a generic modified gravity scenario. In Section~\ref{sec:dataanalysis}, after briefly describing the data used in our analysis and the surveys from which they were obtained, we detail our methodology for cosmological inference, in particular how we modelled the theoretical power spectra for comparison with our data vector. In Section~\ref{sec:results} we present our Horndeski gravity constraints, while the agreement of our $\Lambda$CDM results with the ones obtained by vU18 is shown in Appendix~\ref{app:comparison}. We also describe how our results compare with the \textit{Planck} results, and present constraints obtained from a combined analysis with CMB measurements. Finally, we draw our conclusions in Section~\ref{sec:conclusions}.  

\section{Horndeski theories}\label{sec:Horndeski}

The most general scalar-tensor theory of gravity that is four-dimensional, Lorentz-invariant, local, energy-momentum conserving and has equations of motion with derivatives not higher than second order, can be written as \citep{Horndeski1974}:
\begin{align}\label{eq:Horndeski_action}
S [ g_{\mu \nu}, \phi ] &= \int \mathrm{d}^4 x \sqrt{- g} \left[ \sum_{i=2}^5 \frac{1}{8 \pi G_N} \mathcal{L}_i[g_{\mu \nu}, \phi] + \mathcal{L}_m[g_{\mu \nu}, \psi_m] \right],\\
\mathcal{L}_2 &= G_2 (\phi, X), \nonumber \\
\mathcal{L}_3 &= -G_3(\phi, X) \Box \phi, \nonumber \\
\mathcal{L}_4 &=  G_4 (\phi, X) R + G_{4X}(\phi, X) \left[ (\Box \phi)^2 - \phi_{;\mu\nu} \phi^{;\mu\nu} \right],\nonumber \\
\mathcal{L}_5 &= G_5 (\phi, X) G_{\mu\nu} \phi^{;\mu\nu} \nonumber \\ 
&- \frac{1}{6}G_{5X} (\phi, X) \left[ (\Box \phi)^3 + 2 \phi_{;\mu}{}^{\nu} \phi_{;\nu}{}^{\alpha} \phi_{;\alpha}{}^{\mu} - 3 \phi_{;\mu\nu} \phi^{;\mu\nu} \Box \phi \right]. \nonumber
\end{align}  
where $g=\mathrm{det} \, g_{\mu \nu}$, $R$ is the Ricci scalar and the integration in $\mathrm{d}^4 x$ is carried out over the whole four-dimensional spacetime. The $G_i(\phi, X)$ are arbitrary functions of the additional scalar field $\phi$ and its kinetic term $X = -\frac{1}{2} \nabla _\mu \phi \nabla^{\mu} \phi$. The choice of the $G_i(\phi, X)$ functions specifies the single modified gravity model considered within the Horndeski class. The four contributions $\mathcal{L}_i$ to the gravitational sector depend on the $G_i(\phi, X)$ functions and on their partial derivatives, denoted with subscripts $\phi, X$, e.g. $G_{iX} = \partial G_i / \partial X$. The Horndeski Lagrangian only considers universal coupling between the metric and the matter fields (collectively described by $\psi_m$ and contained in the matter Lagrangian $\mathcal{L}_m$), which are therefore uncoupled to the scalar field. Most of the universally coupled models with one additional scalar degree of freedom belong to the Horndeski class. These include for example quintessence \citep{Ratra1988, Wetterich1988}, Brans-Dicke models \citep{Brans1961}, $k$-essence \citep{Armendariz1999, Armendariz2001}, kinetic gravity braiding \citep{Deffayet2010, Kobayashi2010, Pujolas2011}, covariant galileons \citep{Nicolis2009, Deffayet2009}, disformal and Dirac-Born-Infeld gravity \citep{deRham2010a, Zumalacarregui2013, Bettoni2013}, chameleons \citep{Khoury2004b, Khoury2004a}, symmetrons \citep{Hinterbichler2010, Hinterbichler2011}, Gauss-Bonnet couplings \citep{Ezquiaga2016} and models screening the cosmological constant \citep{Charmousis2012, MartinMoruno2015}, all variants of $f(R)$ \citep{Carroll2004} and $f(G)$ \citep{Carroll2005} theories.

We now turn to the description of linear cosmological perturbations in Horndeski gravity. Considering linear scalar perturbations on a Friedmann-Robertson-Walker metric, assuming spatial flatness on large scales one can write the line element in Newtonian gauge, using cosmic time $t$ and comoving coordinates $\mathbf{x}$, as
\begin{align}
\label{def:pot}
\mathrm{d}s^2 = -\left(1+2\frac{\Phi}{c^2} \right) c^2 \mathrm{d}t^2 + a^2 \left( t \right) \left(1-2\frac{\Psi}{c^2} \right) \mathrm{d} \mathbf{x}^2,
\end{align}
with the Bardeen potentials $\Phi$ and $\Psi$ satisfying the condition $\Phi = \Psi$ in General Relativity in absence of anisotropic stress; for a modified gravity theory, this equality in general does not hold. 

We fix the background expansion to $\Lambda$CDM and study the evolution of linear cosmological perturbations in Horndeski gravity. It has been shown that linear perturbations in Horndeski theories can be parameterized by means of four functions of (conformal) time $\tau$ only \citep{Gleyzes2013, BelliniSawicki2014}, which purely affect structure formation leaving the background expansion unchanged. A possible choice for these four functions is particularly informative on the physical meaning associated to each of them. We will collectively denote these four specific functions as $\alpha(\tau)$ and briefly comment on their physical interpretation here, referring to \citet{BelliniSawicki2014} and references therein for a more complete description:
\begin{itemize}
\item $\alpha_K$ is the \textit{kineticity} term, i.e. the kinetic energy of the scalar perturbations arising directly from the action. In the quasi-static approximation where time derivatives are negligible compared to space derivatives, $\alpha_K$ does not enter the equations of motion and is therefore largely unconstrained by cosmic shear and other large-scale structure probes (\citealp{Bellini2016, Alonso2017, SpurioMancini2018, Reischke2018}). In our analysis we will fix $\alpha_K$ to zero, i.e. its General Relativity value. This does not affect the constraints on the other $\alpha$ functions, since $\alpha_K$ is largely uncorrelated with all other parameters \citep{SpurioMancini2018, Reischke2018} \footnote{In practice, when running our MCMC chains we fix $\alpha_K$ to a very small value; for example, when using the parameterization $\alpha_K (\tau) = \hat{\alpha}_K \Omega_{\mathrm{DE}}(\tau)$ (see the end of Section~\ref{sec:Horndeski}), we set $\hat{\alpha}_K = 0.01$. Observationally, the difference between predicted matter power spectra (which enter the expression for the power spectra of the cosmological probes in our analysis) for $\hat{\alpha}_K=0.01$ and $\hat{\alpha}_K=0$ would not be distinguishable, while setting $\hat{\alpha}_K = 0$ produces numerical instabilities in \textsc{HiClass}. For this reason we fix $\hat{\alpha}_K$ to 0.01; we also notice that any other value for $\hat{\alpha}_K$ would be equally acceptable, as $\alpha_K$ is in any case unconstrained by large-scale structure probes, and largely uncorrelated with all other parameters.};

\item $\alpha_B$ is the \textit{braiding} term, which describes mixing of the scalar field with the metric kinetic term, leading to what is typically interpreted as a fifth force between massive particles influencing the structure formation rate. As shown by  \citet{BelliniSawicki2014}, $\alpha_B$ introduces a new scale dependence of the theory beyond the Jeans length. The dark energy component clusters at small scales only if there is braiding, $\alpha_B \neq 0$ \citep{Bellini2016};

\item $\alpha_M$ is the \textit{Planck-mass run rate}, defined by 
\begin{align}
\alpha_M \equiv \frac{\mathrm{d \, ln}M_*^2}{\mathrm{d \, ln} \, a}.
\end{align}
$\alpha_M$ describes the rate of evolution of the effective Planck mass $M_*^2$, defined as the dimensionless product of the normalization of the kinetic term for gravitons and $8 \pi G_N$ measured on Earth, i.e. it describes a time evolution of the effective gravitational constant which modifies the growth of perturbations when there is non-zero braiding. A mere fixed redefinition of the Planck mass would not produce detectable effects on structure formation, as it could always be reabsorbed by an appropriate rescaling of the action. Theories with $\alpha_M \neq 0$ are non-minimally coupled. As shown in \citet{Saltas2014}, $\alpha_M$ generates anisotropic stress and modifies the propagation of gravitational waves; 

\item $\alpha_T$ is the \textit{tensor speed excess,} indicating deviations of the propagation speed of gravitational waves from the speed of light. Recently, very strong constraints have been placed on $\alpha_T$ by the measurement of the gravitational wave speed derived by the detection of the binary neutron star merger GW170817 and the associated gamma ray burst GRB170817A \citep{Abbott2017b, Abbott2017a, Baker2017, Creminelli2017, Ezquiaga2017, Sakstein2017, Lombriser2017, Bettoni2017}. Since the tensor speed has been found to be very close to that of light, the value of $\alpha_T$ at present time has been constrained to be very close to zero. While in principle this constraint may not apply to $\alpha_T$'s value at all times \citep[see e.g.][]{Amendola2018, Battye2018, deRham2018}, setting $\alpha_T$ to vanish across all cosmic history is the simplest choice and in this work we will stick to this option, following a common choice in recent literature \citep{Kreisch2017, SpurioMancini2018, Reischke2018, Denissenya2018}. We also notice that, since we will be mainly interested in a parameterization of the $\alpha$ functions that make them vary linearly with the dark energy density fraction (as discussed later in this Section), for our purposes having constraints that make $\alpha_T$ vanish in our local Universe also imply that $\alpha_T$ vanishes across all cosmic history. 

We remark that the other three $\alpha$ functions are still free to vary, although again, we fix $\alpha_K$ in our analysis because it is unconstrained by large-scale structure probes and uncorrelated with all other parameters. This leaves us with two Horndeski functions to constrain, $\alpha_B$ and $\alpha_M$. For details on the status of studies of Horndeski theories after the constraints on $\alpha_T$ from gravitational waves, we refer the reader to \cite{Ezquiaga2017}, who identify the models within the Horndeski classes that are still viable after GW170817 \citep[see also][]{Kase2018}, and \cite{Peirone2018}, who show that even the strict bound on the present-day gravitational wave speed does not exclude at all nontrivial signatures of modified gravity that can be measured in linear cosmic perturbations. Excluding modifications to $\alpha_T$ (as we assume throughout our analysis, as well as fixing $\alpha_K$ and the expansion history) effectively reduces the large class of Horndeski models to a more limited set of theories; this includes prototypical examples of dark energy and modified gravity theories, such as quintessence, $f(R)$, Brans-Dicke, kinetic gravity braiding models, while more complicated theories such as Galileon models \citep{Ezquiaga2017, Ezquiaga2018} are excluded.
\end{itemize}

These $\alpha(\tau)$ functions are all identically vanishing in $\Lambda$CDM: any clear detection of values different from zero for these functions would mark a deviation from the concordance cosmological model. In this paper, we aim at setting constraints on the $\alpha(\tau)$ functions. To achieve this goal, we first need to choose a parameterization for their time evolution. Our first choice is to consider their time evolution to trace that of the dark energy component, to which they are proportional. We set then:
\begin{align}\label{eq:alpha_time_evolution}
\alpha_i (\tau) = \hat{\alpha}_i \, \Omega_{\mathrm{DE}}(\tau).
\end{align}
The first constraints we present in Section~\ref{sec:results} will be on the proportionality coefficients $\hat{\alpha}_i$. This parameterization is the simplest and most common in the literature \citep[as used e.g. in][]{PlanckCollaboration2016} and can already provide comprehensive information on Horndeski gravity, as remarked by \citet{Gleyzes2017}. 

The second parameterization we consider, also studied in \citet{PlanckCollaboration2016}, represents a subclass of models, specifically k-\emph{essence} conformally coupled to gravity. In this parameterization the four $\alpha(\tau)$ functions can be specified through a single function of time $\Omega\left(\tau\right)$ affecting three of the $\alpha(\tau)$ functions:
\begin{align}
\begin{cases}
\alpha_{\textrm{K}}= & \frac{3\left(\rho_{\textrm{DE}}+p_{\textrm{DE}}\right)}{H^{2}}+\frac{3\Omega\left(\rho_{\textrm{m}}+p_{\textrm{m}}\right)}{H^{2}\left(1+\Omega\right)}-\frac{\Omega^{\prime\prime}-2aH\Omega^{\prime}}{a^{2}H^{2}\left(1+\Omega\right)}\\
\alpha_{\textrm{M}}= & \frac{\Omega^{\prime}}{aH\left(1+\Omega\right)}\\
\alpha_{\textrm{B}}= & -\alpha_{\textrm{M}}\\
\alpha_{\textrm{T}}= & 0\,,
\end{cases}\label{eq:planck_parametrization}
\end{align}
where a prime denotes derivative with respect to conformal time. The parameterization for $\Omega(\tau)$ is given by $\Omega_0 \, a(\tau)$: the only free modified gravity parameter becomes then $\Omega_0$.

\section{Theoretical power spectra}\label{sec:theory}

In this section we summarise the mathematical expressions for the projected power spectra of the three cosmological probes considered in our analysis. Our data vector, which we share with vU18, is composed of estimates of these power spectra, as functions of the tomographic bin and the angular multipole $\ell$. There are multiple advantages in using estimators of power spectra, acting in Fourier space, over real space correlation functions, although the latter may be easier to measure \citep{Kohlinger2016}: we mention here in particular the fact that the covariance matrix of the power spectra is more diagonal than its real-space counterpart, which contributes to a cleaner separation of scales; furthermore, power spectrum estimators can be used to extract the $B$-mode part of the signal, which in cosmic shear studies serves as a systematic check because, in absence of systematics, it should be consistent with zero at lowest order. 

For estimation of the power spectra the authors of vU18 followed the formalism originally developed for cosmic shear by \citet{Schneider2002} (and extended by vU18 to galaxy-galaxy lensing and galaxy clustering) based on simple integrals with appropriate weighting functions over the real-space correlation functions, which can be measured with existing public code such as \textsc{TreeCorr}\footnote{\url{https://github.com/rmjarvis/TreeCorr}} \citep{Jarvis2004}, employed by vU18; we refer the reader to vU18 for details.

We conform to vU18 in referring to the power spectra associated to cosmic shear, galaxy-galaxy lensing and galaxy clustering as $P^{\mathrm{E}}, P^{\mathrm{gm}}, P^{\mathrm{gg}}$, respectively. In the cosmic shear case, the label stands for $E$-mode. vU18 also measured the $B$-mode power spectra and found a tentative signal; however, based on the assumption that a potential systematic affects $E$- and $B$-modes in the same way, they created a new data vector where they substracted the B-modes from the E-modes, and found that this correction shifted their main cosmological inference result by less than $0.5\sigma$. Since $P^{E}$ does not vary rapidly with $\ell$, vU18 only needed a few $\ell$-bins to capture most of the cosmological information. They used five logarithmically-spaced bins, whose logarithmic means range from $\ell = 200$ to $\ell = 1500$. $P^{\mathrm{gm}}$ and $P^{\mathrm{gg}}$ were estimated adopting the same $\ell$ ranges. Our data vector is unchanged with respect to the one used by vU18 and as such shares all the aforementioned properties. On the theoretical side, however, differently from vU18 we are working on constraints on Horndeski gravity: therefore, we need to present theoretical predictions for the power spectra in a modified gravity context, to be compared with the estimated power spectra in the cosmological inference process.

\subsection{Modifications of Poisson equation and ratio of Bardeen potentials}

In order to provide equations for the power spectra of the three probes in a general modified gravity scenario, we need to introduce the modifications to the Poisson equation and the ratio of the Bardeen potentials that distinguish a modified gravity theory from General Relativity in absence of anisotropic stress. In Newtonian gauge the linear perturbation equations in Fourier space in a general modified gravity scenario are given by \citep{Saltas2014, Zhang2007, Amendola2008}
\begin{align}\label{eq:mu}
\Phi &= - \frac{3}{2} \frac{\Omega_m H_0^2}{k^2} \frac{\delta}{a} \; \mu (k, \chi) \, ;\\
\frac{\Psi}{\Phi} &= \eta (k, \chi) \label{eq:eta} \, ,
\end{align}
where $\mu$ and $\eta$ are in general functions of both time and scale, and equal to unity in General Relativity in absence of anisotropic stress. In the quasi-static limit, where one neglects time derivatives in the Einstein equations for the perturbations \citep{Sawicki2015}, an explicit formula can be found, connecting the $\alpha(\tau)$ functions directly to the $\mu$ and $\eta$ functions \citep[see e.g. Appendix A in][]{Alonso2017}. However, the validity of the quasi-static approximation depends on the specific modified gravity model considered and the range of scales probed by the cosmological survey considered in the analysis \citep{Baker2015}. We stress here that we do not make use of the quasi-static approximation: as described in Section~\ref{sec:dataanalysis}, we source the potential and density statistics directly from the Boltzmann code \textsc{HiClass}, which does not assume the quasi-static regime. The values of the $\alpha(\tau)$ functions at different times is a direct output of the code, while the values of $\mu$ and $\eta$ can be easily derived from the evolved potential and density statistics \citep[as done e.g. in][]{SpurioMancini2018, Reischke2018}. 

The equations presented below are for this general modified gravity scenario, and the ones used in vU18 for $\Lambda$CDM can be deduced from ours by setting $\mu = \eta = 1$ identically, which is equivalent to setting all $\alpha(\tau)$ functions to zero. For all our power spectrum estimators we assume the extended Limber approximation, in the form given by \citet{Loverde2008}, which uses $\ell+1/2$ in the argument of the matter power spectrum but no additional prefactors. Many recent papers have demonstrated that for cosmic shear these approximations are excellent on the scales that we consider \citep{Kitching2016,Lemos2017,Kilbinger2017}, and this is also true for galaxy-galaxy lensing and galaxy clustering power spectra (see vU18). A detailed derivation of the expression for the cosmic shear power spectrum is reported in Appendix~\ref{app:cosmicshear_mg_power}; expressions for galaxy-galaxy lensing and galaxy clustering power spectra can be derived following the same approach.

\subsection{Cosmic shear}\label{sec:cosmicshear}

The weak lensing convergence power spectrum can be obtained from the 3D matter power spectrum $P_\delta$ via
\begin{align}\label{eq:pe}
P _{i j}^{E} (\ell) &= \left( \frac{3 H_0^2 \Omega_m}{2 c^2} \right)^2 \int_0^{\chi_H} \, \mathrm{d} \chi \, \frac{g_i(\chi) g_j(\chi)}{a(\chi)^2} \, P_\delta \, \left(\frac{\ell + 1/2}{\chi};{\chi} \right) \\ &  \times \, {\mu\left(\frac{\ell +1/2}{\chi}, \chi\right)^2 \, \left(\frac{1 + \eta\left(\frac{\ell+1/2}{\chi}, \chi\right)}{2}\right)^2}, \nonumber
\end{align}
where indices $i, j$ label the tomographic bins, $H_0$ is the Hubble constant, $\Omega_m$ the present-day matter density parameter, $c$ the speed of light, $\chi$ the comoving distance (we assume spatial flatness here and throughout the paper), $a(\chi)$ the scale-factor, $\chi_{\rm H}$ the Hubble radius today, and $g_i(\chi)$ a geometric weight factor, which depends on the source redshift distribution $n_i(z)$ for tomographic bin $i$, $n_i(z) \, {\rm d}z=n_i(\chi) \, {\rm d}\chi$:
\begin{align}\label{eq:g_i}
g_i(\chi) &= \int_\chi^{\chi_H} \, \mathrm{d} \chi' \, n_i(\chi')  \, \frac{\chi'-\chi}{\chi'}.
\end{align}
The $n_i(\chi)$ distributions are normalised such that $\int \, \mathrm{d}\chi \, n_i(\chi) = 1 \, \forall i$.

\subsection{Galaxy-galaxy lensing}\label{sec:gm}

The cross-correlation spectrum between lensing convergence and matter density has the following form 
\begin{align}\label{eq:pgm}
P^{\mathrm{gm}}_{ij} (\ell) & = b_{i} \left( \frac{3 H_0^2 \Omega_m}{2 c^2} \right) \int_0^{\chi_H} \, \mathrm{d} \chi \, \frac{n_{Fi}(\chi) g_j(\chi)}{a(\chi) \chi} P_\delta \left(\frac{\ell + 1/2}{\chi};{\chi} \right) \\ & \times \,   \mu\left(\frac{\ell+1/2}{\chi}, \chi\right)^2 \left(\frac{1 + {\eta}\left(\frac{\ell+1/2}{\chi}, \chi\right)}{2}\right), \nonumber
\end{align}
with $n_{\rm F_i}(\chi)$ the redshift distribution of the foreground sample labelled by index $i$ (here indices $i$ and $j$ run over foreground samples and cosmic shear tomographic bins, respectively). Similarly to vU18, we use an effective linear bias $b_i$ for each foreground sample (see Section~\ref{sec:alignments} for details). 

\subsection{Galaxy clustering}\label{sec:gg}

The angular power spectrum can be determined from the matter power spectrum via
\begin{align}\label{eq:pgg}
P^{\mathrm{gg}}_{ii} (\ell) = b_{i} b_{i} \int_0^{\chi_H} \, \mathrm{d} \chi \, \frac{n_{Fi}(\chi) n_{Fi}(\chi)}{\chi^2} P_\delta \left(\frac{\ell + 1/2}{\chi};{\chi} \right)  \, \mu\left(\frac{\ell+1/2}{\chi}, \chi\right)^2,
\end{align}
where, as above, $b_i$ corresponds to the effective bias of the sample and, as in vU18, we consider only auto-correlations of the tomographic bin power spectra. Our dataset consists of two GAMA foreground samples (see Section~\ref{sec:data}), one at lower and the other at higher redshift. Following \citet{Loverde2008}, for our two foreground samples we obtain threshold scales for the validity of the Limber approximation of $\ell\gtrsim15$ and $\ell\gtrsim25$, respectively.~Since the minimum $\ell$ scale entering the analysis is 150, the Limber approximation is valid for both our samples at subpercent level \citep[see also][]{Kilbinger2017}. 

\begin{table}
	\centering
	\caption{Priors on the fit parameters. Rows 1--5 contain the priors on cosmological parameters, rows 6--9 the priors on astrophysical `nuisance' parameters, rows 10--13 the priors on modified gravity parameters. All prior distributions are uniform within their ranges.}
	\begin{tabular}{l l c} 
		\hline
		Parameter & Description &  Prior range \\
		\hline\hline
		$\omega_{\rm cdm}$ & Cold dark matter density & $[0.01,0.99]$ \\
		$\omega_{\rm b}$ & Baryon density & $[0.019,0.026]$ \\
		$\ln(10^{10}A_{\rm s})$ & Scalar spectrum amplitude & $[1.7,5.0]$ \\
		$n_{\rm s}$ & Scalar spectral index & $[0.7,1.3]$ \\
		$h$ & Dimensionless Hubble parameter & $[0.64,0.82]$ \\
		\hline
		$A_{\rm IA}$ & Intrinsic alignment amplitude & $[-6,6]$ \\
		$c_{\mathrm{min}}$ & Baryonic feedback amplitude & $[2,4]$ \\
		$\mathrm{bias}_{\rm z1}$ & Galaxy bias of low-$z$ lens sample & $[0.1,5]$ \\
		$\mathrm{bias}_{\rm z2}$ & Galaxy bias of high-$z$ lens sample & $[0.1,5]$ \\
		\hline
		$\hat{\alpha}_{\mathrm{B}}$ & Prop. coeff. $\alpha_{\mathrm{B}}(\tau) = \hat{\alpha}_{\mathrm{B}} \Omega_{\mathrm{DE}}(\tau)$& $[-2,2]$\\
		$\hat{\alpha}_{\mathrm{M}}$ & Prop. coeff. $\alpha_{\mathrm{M}}(\tau) = \hat{\alpha}_{\mathrm{M}} \Omega_{\mathrm{DE}}(\tau)$ &$[-2,2]$\\
		$\Omega_0$ & Prop. coeff. $\Omega(\tau) = \Omega_0 a(\tau)$ & $[0,1]$ \\
		log $k_s$ & screening scale & $[-1, 1]$\\
		\hline
	\end{tabular}
	\label{tab_prior}
\end{table} 

\section{Data analysis}\label{sec:dataanalysis}

\begin{table*}
	\begin{center}
		\begin{tabular}{|c|c|c|c|c|c|c|c|}
			\hline
			\hline
			Parameter          &  $P^{\mathrm{E}}$        & $P^{\mathrm{E}} + P^{\mathrm{gm}} + P^{\mathrm{gg}}$ & $P^{\mathrm{E}} + P^{\mathrm{gm}}$ & $P^{\mathrm{E}} + P^{\mathrm{gg}}$ & $P^{\mathrm{gm}} + P^{\mathrm{gg}}$ & \textit{Planck} & $P^{\mathrm{E}} + P^{\mathrm{gm}} + P^{\mathrm{gg}} +$ \textit{Planck} \\
			\hline
			\hline  
			$\omega_{\mathrm{cdm}}$                    &   $0.119_{-0.025}^{+0.015}$   &  $0.121_{-0.018}^{+0.016}$    &   $0.118_{-0.020}^{+0.015}$      &   $0.119_{-0.017}^{+0.013}$  &   $0.131_{-0.022}^{+0.017}$ & $0.117_{-0.002}^{0.002}$ & $0.117_{-0.002}^{+0.002}$\\[5pt]
			$\mathrm{ln}10^{10}A_s$                       &   $3.01_{-0.25}^{+0.30}$   &  $3.02_{-0.19}^{+0.24}$    &   $3.10_{-0.26}^{+0.24}$      &   $3.01_{-0.18}^{+0.29}$  &   $3.06_{-0.24}^{+0.25}$ & $3.12_{-0.01}^{0.01}$ & $3.12_{-0.01}^{+0.01}$ \\[5pt]
			$\omega_b$                                &   $0.0225_{-0.0037}^{+0.0035}$   &  $0.0224_{-0.0037}^{+0.0038}$    &   $0.0221_{-0.0033}^{+0.0034}$      &   $0.0223_{-0.0036}^{+0.0038}$  &   $0.0226_{-0.0038}^{+0.0037}$ & $0.0225_{0.0003}^{0.0003}$ & $0.0225_{-0.0002}^{+0.0002}$ \\[5pt]
			$n_s$                                         &   $1.07_{-0.10}^{+0.23}$   &  $1.13_{-0.05}^{+0.17}$    &   $1.13_{-0.07}^{+0.17}$      &   $1.14_{-0.05}^{+0.16}$  &   $1.08_{-0.09}^{+0.22}$ & $0.97_{-0.01}^{+0.01}$ & $0.97_{-0.01}^{+0.01}$ \\[5pt]
			$h$                                           &   $0.73_{-0.09}^{+0.09}$   &  $0.75_{-0.03}^{+0.07}$    &   $0.77_{-0.03}^{+0.05}$      &   $0.77_{-0.02}^{+0.05}$  &   $0.75_{-0.03}^{+0.07}$ & $0.69_{-0.01}^{+0.01}$ & $0.69_{-0.01}^{+0.01}$ \\[5pt]
			$c_{\mathrm{min}}$                            &   $3.20_{-0.46}^{+0.80}$   &  $3.34_{-0.37}^{+0.66}$    &   $3.28_{-0.31}^{+0.72}$      &   $3.27_{-0.29}^{+0.73}$  &   $3.16_{-0.48}^{+0.84}$ & - & $3.69_{-0.08}^{+0.31}$\\[5pt]
			$A_{\mathrm{IA}}$                             &   $0.44_{-0.68}^{+1.16}$   &  $1.37_{-0.44}^{+0.48}$    &   $1.39_{-0.46}^{+0.51}$      &   $0.84_{-0.61}^{+0.88}$  &   $1.51_{-0.54}^{+0.51}$ & - & $1.40_{-0.35}^{+0.38}$\\[5pt]
			$\mathrm{bias}_{z1}$                          &             -                    &  $1.01_{-0.08}^{+0.06}$    &   $0.99_{-0.09}^{+0.09}$      &   $1.01_{-0.08}^{+0.05}$  &   $1.01_{-0.09}^{+0.06}$ & - & $1.02_{-0.06}^{+0.04}$\\[5pt]
			$\mathrm{bias}_{z2}$                          &             -                    &  $1.05_{-0.09}^{+0.05}$    &   $1.06_{-0.09}^{+0.06}$      &   $1.06_{-0.10}^{+0.05}$  &   $1.05_{-0.09}^{+0.06}$ & - & $1.05_{-0.06}^{+0.03}$\\[5pt]
			$\sigma_8$                                    &   $0.86_{-0.08}^{+0.07}$   &  $0.90_{-0.05}^{+0.06}$    &   $0.93_{-0.06}^{+0.07}$      &   $0.90_{-0.05}^{+0.06}$  &   $0.95_{-0.07}^{+0.07}$ & $0.87_{-0.02}^{+0.01}$ & $0.86_{-0.01}^{+0.01}$\\[5pt]
			$\Omega_m$                                    &   $0.27_{-0.04}^{+0.03}$   &  $0.26_{-0.02}^{+0.02}$    &   $0.24_{-0.03}^{+0.02}$      &   $0.24_{-0.03}^{+0.02}$  &   $0.28_{-0.03}^{+0.02}$ & $0.29_{-0.01}^{+0.01}$& $0.29_{-0.01}^{+0.01}$\\[5pt]
			$S_8$                                         &   $0.803_{-0.051}^{+0.054}$   &  $0.835_{-0.039}^{+0.041}$    &   $0.830_{-0.042}^{+0.053}$      &   $0.806_{-0.044}^{+0.047}$  &   $0.905_{-0.060}^{+0.057}$ & $0.859_{-0.032}^{+0.029}$ & $0.843_{-0.024}^{+0.021}$ \\[5pt]
			$\hat{\alpha}_{\mathrm{B}}$                   &   $0.05_{-0.36}^{+0.30}$   &  $0.20_{-0.33}^{+0.20}$    &   $0.26_{-0.44}^{+0.33}$      &   $0.23_{-0.33}^{+0.20}$  &   $0.28_{-0.45}^{+0.22}$ & $0.79_{-0.71}^{+0.42}$ & $0.36_{-0.22}^{+0.18}$\\[5pt]
			$\hat{\alpha}_{\mathrm{M}}$                   &   $0.46_{-0.61}^{+0.25}$   &  $0.25_{-0.29}^{+0.19}$    &   $0.31_{-0.44}^{+0.26}$      &   $0.23_{-0.30}^{+0.20}$  &   $0.36_{-0.44}^{+0.24}$ & $0.18_{-0.52}^{+0.35}$ & $0.15_{-0.31}^{+0.13}$\\[5pt]
			$\mathrm{log} \, k_s$                                         &   $0.00_{-1.00}^{+1.00}$   &  $-0.02_{-0.98}^{+0.74}$    &   $-0.01_{-0.99}^{+1.01}$      &   $0.01_{-1.01}^{+0.99}$  &   $0.07_{-1.06}^{+0.87}$ & - & $0.11_{-1.11}^{+0.89}$\\[5pt]
			\hline
			\hline
			$\chi^2$                                      &  63.2                   & 
			115.7                  &   100.1                     &   67.9                 &
			42.8 & 10697 & 10821 \\
			d.o.f.                                        &  40                        &  88            &   78                          &   48                    & 38 &  & \\[5pt]                
			\hline
		\end{tabular}
	\end{center}
	\caption{Mean and marginalised 68 per cent credibility interval on the parameters listed obtained in Horndeski gravity with all different combinations of probes. The parameterization for the $\alpha$ functions is given by proportionality to $\Omega_{\mathrm{DE}}$ so that the free parameters become the proportionality coefficients (in this case $\hat{\alpha}_{\mathrm{B}}$ and $\hat{\alpha}_{\mathrm{M}}$, since we fix $\alpha_{\mathrm{T}} = \alpha_{\mathrm{K}} = 0$). The nuisance parameters (discussed in Secs.~\ref{sec:nonlinear},~\ref{sec:screening} and \ref{sec:alignments}) are the \textsc{HMcode} parameter $c_{\mathrm{min}}$, the effective bias values $\mathrm{bias}_{z1}$ and $\mathrm{bias}_{z2}$ for the low-$z$ and high-$z$ foreground samples, respectively, the intrinsic alignment amplitude $A_{\mathrm{IA}}$ and the screening scale $k_s$. See Sec.~\ref{sec:comparison_Planck} for a discussion of the $\chi^2$ and number of degrees of freedom (d.o.f.) when \textit{Planck} data are considered in the analysis.}\label{tab:ResultsMGproptoOmega}
\end{table*}

\begin{table*}
	\begin{center}
		\begin{tabular}{|c|c|c|c|c|c|}
			\hline
			\hline
			Parameter                           &  $P^{\mathrm{E}}$      & $P^{\mathrm{E}} + P^{\mathrm{gm}} + P^{\mathrm{gg}}$ & $P^{\mathrm{E}} + P^{\mathrm{gm}}$ & $P^{\mathrm{E}} + P^{\mathrm{gg}}$ & $P^{\mathrm{gm}} + P^{\mathrm{gg}}$ \\
			\hline
			\hline 
			$\omega_{\mathrm{cdm}}$             &    $0.163_{-0.052}^{+0.037}$     &   $0.138_{-0.026}^{+0.017}$     &  $0.135_{-0.030}^{+0.021}$     &  $0.136_{-0.026}^{+0.021}$     &  $0.151_{-0.033}^{+0.023}$ \\[5pt]
			$\mathrm{ln}10^{10}\mathrm{A_s}$    &    $3.00_{-0.60}^{+0.56}$        &   $3.06_{-0.33}^{+0.34}$        &  $3.21_{-0.43}^{+0.36}$        &  $3.06_{-0.41}^{+0.35}$        &  $2.93_{-0.32}^{+0.42}$ \\[5pt]
			$\omega_b$                          &    $0.0225_{-0.0037}^{+0.0038}$  &   $0.0225_{-0.0035}^{+0.0038}$  & $0.0224_{-0.0036}^{+0.0038}$   &  $0.0223_{-0.0036}^{+0.0035}$  &  $0.0224_{-0.0037}^{+0.0038}$ \\[5pt]
			$n_s$                               &    $1.03_{-0.08}^{+0.27}$        &   $1.10_{-0.05}^{+0.20}$        & $1.05_{-0.10}^{+0.25}$         &  $1.09_{-0.06}^{+0.21}$        &  $1.05_{-0.10}^{+0.25}$         \\[5pt]
			$h$                                 &    $0.73_{-0.09}^{+0.08}$        &   $0.76_{-0.02}^{+0.07}$        & $0.76_{-0.02}^{+0.06}$         &  $0.77_{-0.02}^{+0.05}$        &  $0.76_{-0.02}^{+0.06}$        \\[5pt]
			$c_{\mathrm{min}}$                  &    $3.07_{-0.35}^{+0.93}$        &   $3.10_{-0.38}^{+0.89}$        & $3.21_{-0.26}^{+0.79}$         &  $3.12_{-0.47}^{+0.88}$        &  $3.21_{-0.26}^{+0.79}$         \\[5pt]
			$A_{\mathrm{IA}}$                   &    $0.63_{-0.60}^{1.03}$         &   $1.31_{-0.39}^{+0.43}$        & $1.35_{-0.41}^{+0.45}$         &  $0.70_{-0.61}^{+0.90}$        &  $1.35_{-0.49}^{+0.43}$        \\[5pt]
			$\mathrm{bias}_{z1}$                &               -                  &   $1.03_{-0.11}^{+0.07}$        & $0.97_{-0.15}^{+0.14}$         &  $1.04_{-0.11}^{+0.07}$        &  $1.07_{-0.16}^{+0.09}$\\[5pt]
			$\mathrm{bias}_{z2}$                &               -                  &   $1.10_{-0.13}^{+0.07}$        & $1.15_{-0.17}^{+0.09}$         &  $1.13_{-0.15}^{+0.08}$        &  $1.10_{-0.16}^{+0.10}$ \\[5pt]
			$\sigma_8$                          &    $0.87_{-0.14}^{+0.11}$        &   $0.93_{-0.09}^{+0.08}$        & $0.93_{-0.10}^{+0.08}$         &  $0.92_{-0.10}^{+0.08}$        &  $0.85_{-0.07}^{+0.09}$\\[5pt]
			$\Omega_{\mathrm{m}}$               &    $0.35_{-0.09}^{+0.07}$        &   $0.28_{-0.03}^{+0.02}$        & $0.27_{-0.04}^{+0.03}$         &  $0.27_{-0.04}^{+0.03}$        &  $0.31_{-0.05}^{+0.03}$\\[5pt]
			$S_8$                               &    $0.917_{-0.106}^{+0.096}$     &   $0.895_{-0.074}^{+0.048}$     & $0.879_{ -0.066}^{+0.047}$     & $0.866_{-0.077}^{+0.051}$      & $0.863_{-0.043}^{+0.047}$\\[5pt]
			$\Omega_0$                          &    $0.54_{-0.29}^{+0.46}$        &   $0.21_{-0.21}^{+0.04}$        & $0.31_{-0.31}^{+0.07}$         &  $0.23_{-0.23}^{+0.04}$        &  $0.44_{-0.44}^{+0.56}$ \\[5pt]
			log $k_s$                           &    $0.18_{-0.37}^{+0.48}$        &   $0.05_{-1.05}^{+0.57}$        &  $-0.24_{-0.76}^{+0.28}$       &  $0.01_{-1.01}^{+0.52}$       &  $0.06_{-1.07}^{+0.86}$\\[5pt]
			\hline
			\hline
			$\chi^2$                                      &  62.1                   & 
			116.2                  &   99.3                     &   68.8                 &
			43.5 \\
			d.o.f.                                        &  41                        &  89                        &   79                          &   49                    & 39 \\[5pt]                
			\hline
			
		\end{tabular}
	\end{center}
	\caption{Mean and marginalised 68 credibility interval on the parameters listed obtained in Horndeski gravity with all different combinations of probes. The parameterization for the $\alpha$ functions is the \texttt{planck\_linear} parameterization implemented in \textsc{HiClass}, where the $\alpha$ functions depend on only one function of time $\Omega(\tau)$ (see equations~\ref{eq:planck_parametrization}), which is assumed to be proportional to the scale factor, $\Omega(\tau)=\Omega_0 a(\tau)$. The only free parameter as far as the $\alpha$ functions are concerned becomes then $\Omega_0$. The nuisance parameters are the same as in Table \ref{tab:ResultsMGproptoOmega}.}\label{tab:ResultsMGLinear}
\end{table*}

\begin{figure*}
	\centering
	\includegraphics[width = 0.9\textwidth]{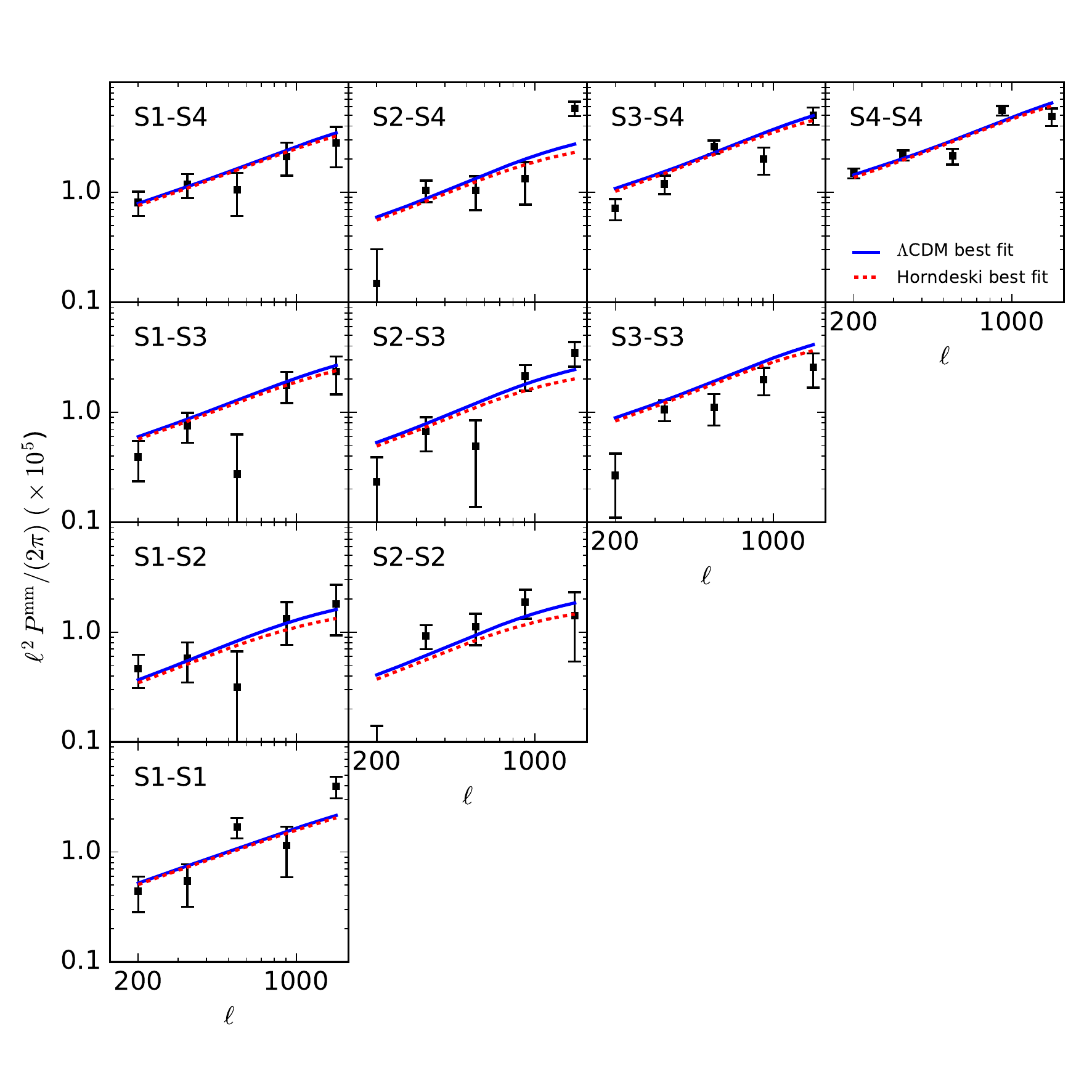}	
	\caption{Theoretical predictions for the cosmic shear power spectra, given the parameter values corresponding to the best fit for the combined cosmic shear - galaxy-galaxy lensing - galaxy clustering analysis, assuming $\Lambda$CDM (\textit{blue}) and Horndeski gravity (\textit{red}), overplotted with the measured values for the power spectra from the KiDS survey (\textit{blue}). Error bars are computed analytically from the covariance matrix. The numbers in each panel indicate which shape (S) samples are correlated.}\label{fig:bestfitpe}
\end{figure*}

\begin{figure*}
		\includegraphics[width=0.9\textwidth]{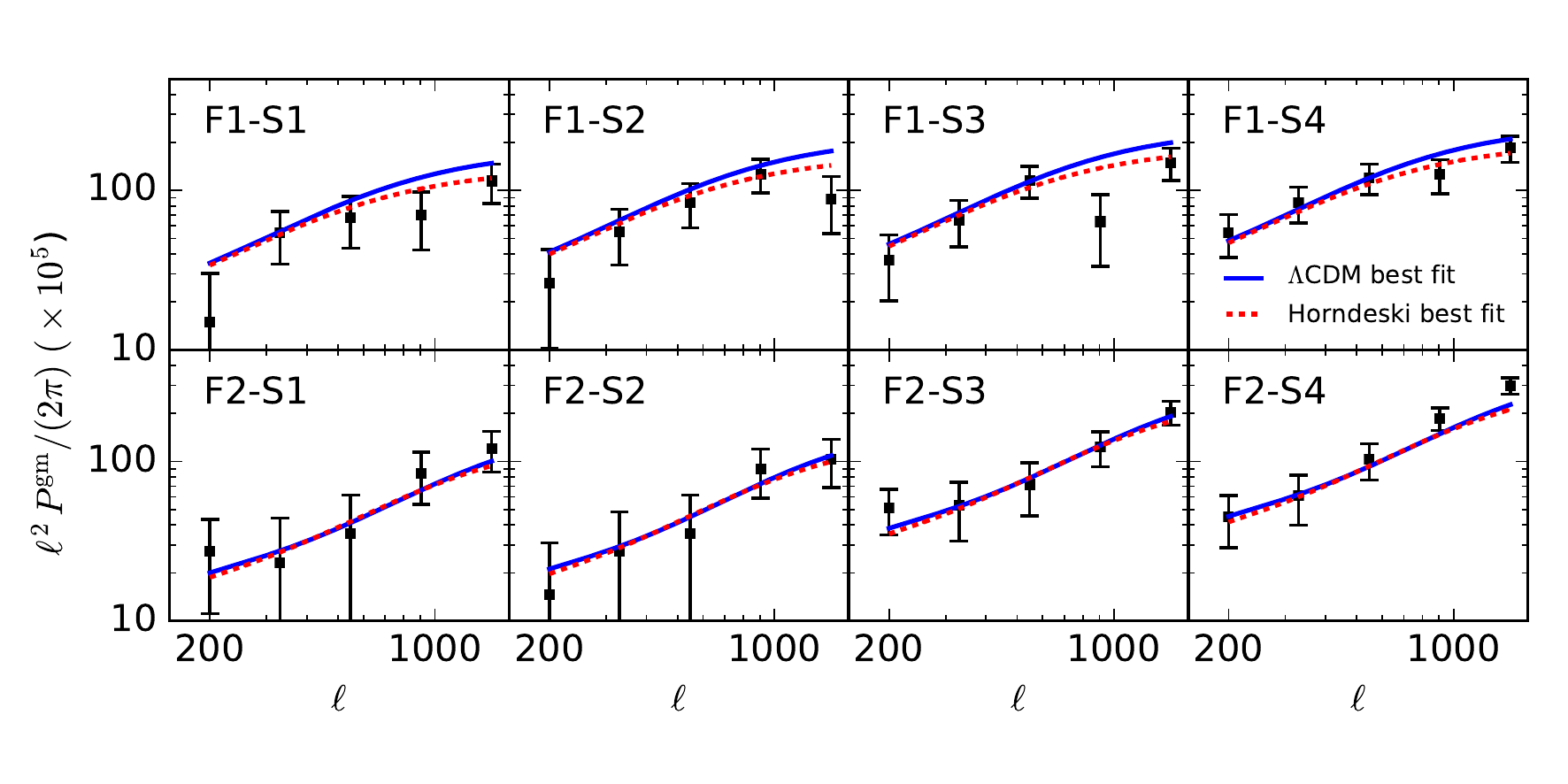}\par 
		\caption{Same as in Fig.~\ref{fig:bestfitpe}, but for galaxy-galaxy lensing power spectra. The numbers in each panel indicate which foreground (F) and shape (S) samples are correlated.}\label{fig:bestfitpgm}
\end{figure*}
\begin{figure}
\centering
		\includegraphics[width=0.9\columnwidth]{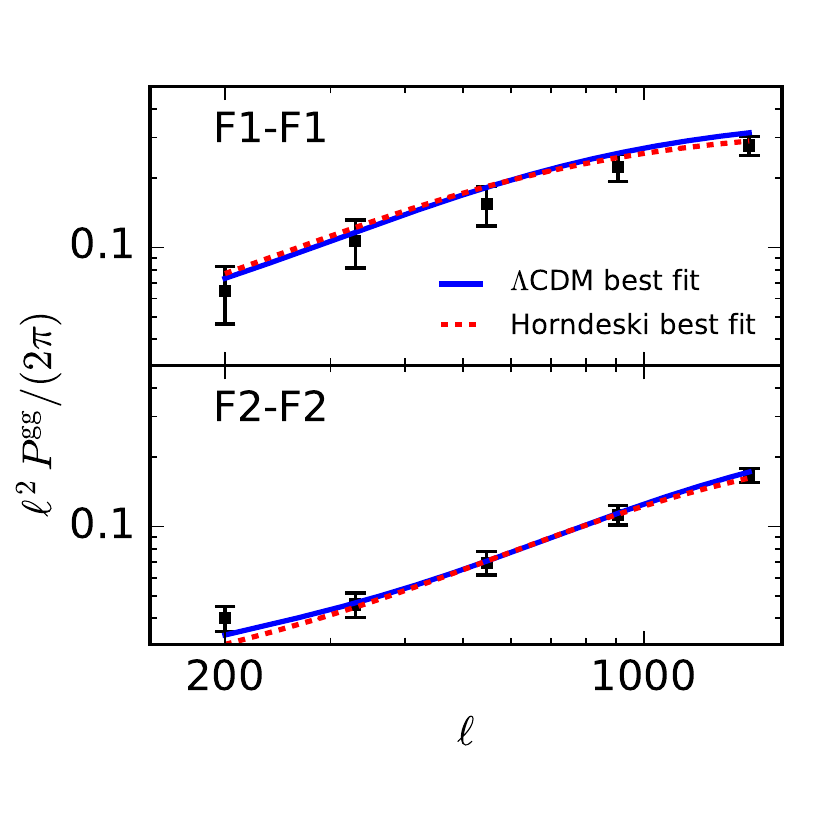}\par
		\caption{Same as in Fig.~\ref{fig:bestfitpe}, but for galaxy clustering power spectra. The numbers in each panel indicate which foreground (F) samples are correlated.}\label{fig:bestfitpgg} 
\end{figure}

In order to obtain constraints on both standard cosmological parameters and those that describe the evolution of linear perturbations in Horndeski gravity (introduced in Section~\ref{sec:Horndeski}), we carry out cosmological inference in a Bayesian framework, sampling the likelihood
\begin{equation} \label{eq:shear_lkl}
-2 \ln {\mathcal{L}(\bm{\theta})} = \sum_{\alpha, \, \beta} d_\alpha(\bm{\theta}) (\mathbf {C}^{-1})_{\alpha \beta} \, d_\beta(\bm{\theta}) \, ,
\end{equation}
where the indices $\alpha$, $\beta$ run over the probes considered, as well as the tomographic bins and the angular multipole $\ell$. The analytical covariance matrix $\mathbf {C}$ is calculated as outlined below in Section~\ref{sec:cov}. 
Equation~(\ref{eq:shear_lkl}) assumes that the estimated power spectra are Gaussian distributed around their mean; it must be noted that this is an approximation, rigorously valid only at high $\ell$ as a consequence of the Central Limit theorem. For the multipole range considered here, this approximation is safe.
The components of the vector multiplying the covariance matrix are calculated as
\begin{equation}
d_\alpha(\bm{\theta}) = P_\alpha - \langle P_\alpha(\bm{\theta}) \rangle^{\rm model} \, ,
\end{equation}
where the dependence on cosmological parameters enters only in the calculation of the predicted power spectra, $\langle P_{\alpha}(\bm{\theta}) \rangle^{\rm model}$, which is carried out according to Eqs.~\ref{eq:pe}, \ref{eq:pgm}, \ref{eq:pgg}, while $P_{\alpha}$ denotes the measured power spectrum, with index $\alpha$ running over the cosmological probe and pair of tomographic bins considered. 

To constrain the cosmological parameters, we use the sampler {\sc MontePython} \citep{Audren2013} \footnote{Version 2.2.2 from \url{https://github.com/baudren/montepython_public}} and build a likelihood module to analyse the three probes considered. The likelihood module is based on the one developed in \citet{Kohlinger2017} for the KiDS-450 quadratic estimator analysis. Our module has been developed considering as high priority the flexibility to choose different combinations of cosmological probes as well as the possibility to perform the analysis in either a standard $\Lambda$CDM cosmological scenario or an extended modified gravity scenario within the Horndeski framework. These two options are possible as a result of interfacing {\sc MontePython} with \textsc{HiClass}~\citep{Zumalacarregui2016}\footnote{\url{https://github.com/miguelzuma/hi_class_public}}, a Boltzmann solver extending {\sc Class}~\citep{Blas2011}\footnote{Version 2.4.5 from \url{https://github.com/lesgourg/class_public}} to Horndeski models, and thus allowing the user to source the matter power spectrum in either a $\Lambda$CDM or a Horndeski cosmological model, respectively (see Section \ref{sec:mg_results} for details). As a sanity check, we verified that we obtain the same $\Lambda$CDM results when running with {\sc HiCLASS} (with Horndeski corrections switched off) as when running directly with {\sc Class}. Our code implementation is publicly accessible\footnote{Our likelihood module will be made publicly available after acceptance of this paper by the Journal.}. 

In contrast, the analysis of vU18 was carried out only in $\Lambda$CDM with a likelihood module built for the sampler {\sc CosmoMC}~ \citep{Lewis2002} for cosmological parameter estimation. The likelihood code of vU18 was based on the one used in \citet{Joudaki2017} for the fiducial KiDS cosmic shear analysis.

For an efficient evaluation of the likelihood we use two sampling methods, namely \textsc{MultiNest}~\citep{Feroz2009}, a multimodal Nested Sampling \citep{Skilling2006} algorithm included with a Python wrapper \citep[\textsc{PyMultiNest};][]{Buchner2014} in \textsc{MontePython}, and highly parallelised affine invariant sampling through the \textsc{CosmoHammer} suite~\citep{Akeret2012}, embedded in \textsc{MontePython}. \textsc{CosmoHammer} embeds in turn \textsc{emcee}, an implementation by \citet{Foreman2013} of the affine invariant ensemble sampler by \citet{Goodman2010}. We verified that both \textsc{MultiNest} and \textsc{CosmoHammer} sampling methods yield inference results consistent between each other.

\subsection{Covariance}
\label{sec:cov}
As can be seen from equation~(\ref{eq:shear_lkl}), errors on cosmological parameters are determined by the parameter covariance matrix $\mathbf{C}$, or more precisely its inverse, the precision matrix $\mathbf{C}^{-1}$. Developing methods for efficient computation of precision matrices is an area of active research \citep[see e.g.][]{Taylor2014, Sellentin2016, Schafer2016, Reischke2017, Heavens2017, Jeffrey2018}. Our covariance matrix is unchanged with respect to the one used by vU18, i.e. is determined analytically, following a similar formalism as in \citet{Hildebrandt2017}. Here we briefly review the main terms of the calculation, referring to vU18 for further details.

The analytical covariance matrix consists of three terms: (i) a Gaussian term accounting for the Gaussian contribution to sample variance, shape noise, and a mixed noise-sample variance term, estimated following \citet{Joachimi2008}, (ii) an in-survey non-Gaussian term from the connected matter trispectrum, and (iii) a super-sample covariance term. The latter accounts for cosmic variance modes larger than the survey window and coupling to smaller modes within. One of the main advantages of the analytical approach to covariance estimation lies indeed in its ability to better account for super-sample covariance. To compute (ii) and (iii), vU18 followed \citet{Takada2013} and the extension to galaxy-galaxy lensing and galaxy clustering presented e.g. in \citet{Krause2017cosmolike}. 

The publicly available covariance that we employ makes idealistic assumptions about the shot noise contribution which leads to an under-prediction of variance on large scales where the survey geometry becomes relevant. This was recently shown to be the reason behind the relatively large $\chi^2$ values found for previous KiDS and DES analyses \citep{Troxel2018b}. As the posterior was largely unaffected, we can proceed with the available covariance matrix but have to keep the above limitations in mind when interpreting the goodness of fit.

We also notice that, similarly to vU18, we do not consider any parameter dependence for the covariance matrix. vU18 checked that this was choice was safe for their analysis. Here, the covariance matrix is calculated at the same fiducial model chosen by vU18. This is motivated by the fact that we choose a fiducial model for our priors that is given by the same $\Lambda$CDM model chosen by vU18. 
\vspace{-0.7cm}
\subsection{Data}\label{sec:data}
\setlength{\parskip}{0pt}
The Kilo-Degree Survey \citep{deJong2013} is an ongoing optical imaging survey that, once completed, will have spanned 1350 deg$^2$ of the sky in four optical bands, $u$, $g$, $r$ and $i$. Our cosmic shear power spectra \footnote{Measured power spectra and covariance matrices used in this analysis and in vU18 are publicly available at \url{http://kids.strw.leidenuniv.nl/sciencedata.php}}, which we have in common with vU18, are estimated from the KiDS-450 shape measurement catalogues \citep{Hildebrandt2017, deJong2017}, containing shape measurements and photometric redshifts of 450 deg$^2$ of data, with galaxies situated in regions of the sky overlapping with the observational window of GAMA. We use the samples associated with the same four KiDS tomographic source redshift bins adopted in \citet{Hildebrandt2017} and vU18, spanning in redshift $z_{\rm B}$ the intervals 0.1 $<z_{\rm B}\leq$ 0.3, 0.3 $<z_{\rm B}\leq$ 0.5, 0.5 $<z_{\rm B}\leq$ 0.7 and 0.7 $<z_{\rm B}\leq$ 0.9. The main properties of the source samples, such as their average redshift, number density and ellipticity dispersion, can be found in Table 1 of \citet{Hildebrandt2017}. Importantly, we use the same redshift distributions used in vU18 and \citet{Hildebrandt2017}.

The Galaxy And Mass Assembly survey \citep{Driver2009, Driver2011, Liske2015} is a spectroscopic survey of $\sim$240$\,$000 galaxies. vU18 used a subset of $\sim$180$\,$000 galaxies fully overlapping with KiDS.~vU18 selected two GAMA samples, a low redshift sample with $z_{\rm spec}<0.2$, and a high redshift sample with $0.2<z_{\rm spec}<0.5$. These two foreground samples are used to determine the galaxy-matter cross-correlation as well as the clustering power spectra; we adopt both from vU18.

\subsection{Non-linear structure formation and baryonic feedback model}\label{sec:nonlinear}

A proper inclusion of a series of astrophysical systematics and an appropriate modelling of the matter power spectrum on non-linear scales are important to derive accurate cosmological constraints. Here we will describe our choice for the non-linear power spectrum and in particular for the description of baryonic feedback, which modifies the matter power spectrum on small, non-linear scales \citep[e.g.][]{ Semboloni2013}.

The effect of non-linear structure formation and baryonic feedback can now be modelled in \textsc{CLASS} using a module called \textsc{HMcode} (Archidiacono, Brieden \& Lesgourgues \textit{in prep.}), based on the results of \citet{Mead2015}, who account for baryonic effects (in particular, AGN feedback) by modifying parameters that describe the shape of dark matter haloes. This reflects, for example, what happens with AGN and supernova feedback, which make haloes less concentrated by pushing material out of them. In \textsc{HMcode} this is modelled by choosing a functional form for the mass-concentration relation of the type
\begin{align}
c(M,z)=c_{\mathrm{min}}\frac{1+z_{\rm f}}{1+z} \;,
\end{align}
where $z_{\rm f}$ is the formation redshift of a halo, which depends on the halo mass. The free parameter in the fit, $c_{\mathrm{min}}$, modulates the amplitude of this mass-concentration relation. It also sets the amplitude of a `halo bloating' parameter $\eta_0$, which changes the halo profile in a mass dependent way \citep[see equation 26 of][]{Mead2015}. vU18 followed the recommendation of \citet{Mead2015} by fixing $\eta_0 = 1.03 -0.11 c_{\mathrm{min}}$; we follow the same choice. Setting $c_{\mathrm{min}}=3.13$ corresponds to a dark-matter-only model. Marginalizing over $c_{\mathrm{min}}$ allows us to account for the uncertainty on this parameter. This will be done in Section~\ref{sec:results} every time we show contour plots for pairs of parameters (the ones not including $c_{\mathrm{min}}$), as these are marginalised over all other parameters except the two considered in each contour plot.

The problem of non-linear contributions in this modified gravity context is even more exacerbated than in $\Lambda$CDM, since clear prescriptions for modified gravity models on non-linear scales are highly model-dependent and currently not available at the required precision. We decided to produce our non-linear corrections using also in this modified gravity case \textsc{HMcode}, which takes into account baryonic effects. The main obstacle in this sense has been the absence of an implementation of \textsc{HMcode} in \textsc{HiClass}. To solve this problem, we developed our own version of \textsc{HMcode} for \textsc{HiClass}, modifying the one provided to us for \textsc{Class}. We remark here that the use of \textsc{HMcode} is not the most rigorous approach to non-linear corrections in modified gravity, because \textsc{HMcode} is a refined version of \textsc{halofit}~\citep{Smith2003}, which has been developed for a $\Lambda$CDM model. However, while prescriptions for single modified gravity models exist \citep[see e.g.][]{Winther2015}, there is no general one for the non-linear power spectrum for the whole Horndeski class of dark energy and modified gravity models. By choosing to deal with nonlinearities using \textsc{HMcode} we hope to give a flavour of the increased constraining power that comes from non-linear scales \citep[as done e.g. in][for future Stage IV surveys]{Alonso2017, SpurioMancini2018, Reischke2018}, however we warn of the necessity of developing proper formalisms for the non-linear matter power spectrum in Horndeski gravity for the purpose of constraining these gravity models with future surveys. 

Having said that, we notice that given the size of the error bars obtained in our analysis, and the fact that we already marginalise over the baryon feedback parameter, we are not sensitive to small modifications in the non-linear model. To substantiate this statement, in Appendix \ref{app:nonlinearities} we studied in detail the effect of non-linear corrections on our cosmological constraints, by introducing an additional phenomenological nuisance parameter that mimics the effect of different modified gravity models on the non-linear matter power spectrum. Our main finding is that our cosmological constraints are largely insensitive to this additional parameter, corroborating the strength of our analysis against different possible choices for the non-linear matter power spectrum. This will certainly need to be reassessed in future work, where the same analysis could be repeated with larger datasets, more sensitive to varying non-linear prescriptions for different modified gravity models.

\subsection{Screening mechanisms}\label{sec:screening}

A screening mechanism is an important ingredient of modified gravity theories. Acting as a non-linear effect, it screens modifications of gravity at small scales to provide agreement with tests of General Relativity in short scales or high-density environments such as the Solar System \citep[see e.g.][]{Koyama2016}. We model screening mechanisms in a phenomenological way, by applying a scale-dependent filter to the effective Newtonian coupling, the gravitational slip and the linear growth factor:
\begin{equation}
\begin{split}\label{eq:screening}
\mu(k,a)\to & \ 1 + \mu_\mathrm{MG}(k,a)\exp(-(k/k_\mathrm{s})^2) \\
\eta(k,a)\to & \ 1 + \eta_\mathrm{MG}(k,a) \exp(-(k/k_\mathrm{s})^2) \\
D_+(k,a)\to & \ D_\mathrm{+GR}(a) + D_\mathrm{+MG}(k,a) \exp(-(k/k_\mathrm{s})^2).
\end{split}
\end{equation}
This achieves the recovery of General Relativity predictions on scales $k>k_s$. A (conservative) value for $k_s$ is $0.1 \, h \, \mathrm{Mpc}^{-1}$ \citep{Alonso2017, SpurioMancini2018, Reischke2018}; we vary over this scale taking a uniform prior on the logarithm of $k_s$ over the interval $[-1,1]$, thus spanning three orders of magnitude for the screening scale.

\subsection{Intrinsic alignment model}\label{sec:alignments}

\indent Intrinsic alignments affect both the cosmic shear power spectrum and the galaxy-matter power spectrum. For the cosmic shear power spectrum, there are two contributions, the intrinsic-intrinsic (II) and the shear-intrinsic (GI) terms, while the galaxy-matter power spectrum has a galaxy-intrinsic (gI) contribution \citep[e.g.][]{Joachimi2010}. We model the intrinsic alignment power spectrum following the non-linear modification of the linear alignment model \citep{Catelan2001, Hirata2004,Bridle2007,Hirata2010, Blazek2017}:
\begin{eqnarray}
P_{\delta{\rm I}}(k,z) = - A_{\rm IA} C_1 \rho_{\rm crit} \frac{\Omega_{\rm m}}{D_+(k, z)} P_\delta(k,z) \equiv F(k, z) P_\delta(k,z) \;,
\label{eq_pi}
\end{eqnarray}
with $P_\delta(k,z)$ the full non-linear matter power spectrum, $D_+(k, z)$ the linear growth factor, normalised to unity at $z=0$ and in general dependent on scale and redshift in modified gravity, $\rho_{\rm crit}$ the critical density, $C_1=5\times10^{-14}h^{-2}M_\odot^{-1}$Mpc$^3$ a normalization constant chosen such that $C_1 \, \rho_{\rm crit} \approx 0.0134$ \citep{Hirata2004, Bridle2007, Joachimi2011b}, and $A_{\rm IA}$ the overall amplitude, which is a free parameter in our model.

As already mentioned in Secs.\ref{sec:gm} and \ref{sec:gg}, to model $P^{\rm gm}$ and $P^{\rm gg}$, we assume that the galaxy bias is constant and scale-independent, to be interpreted as an effective bias since we also consider non-linear scales. This effective bias is fitted separately for the low-redshift and high-redshift foreground sample. Motivation for not using a scale dependence for the bias comes from combined observations of galaxy-galaxy lensing and galaxy clustering, which found the scale dependence to be small \citep[e.g.][]{Hoekstra2002, Simon2007,Jullo2012,Cacciato2012}. We refer to vU18 for details on how this approximation of scale-independent bias is benign in their analysis as well as in ours. For cosmic shear, in a modified gravity context the intrinsic alignments contributions are given by ($i \leftrightarrow j$ stands for the symmetric term obtained swapping indices $i$ and $j$):

\begin{align}
P _{i j}^{GI} (\ell) &= \left( \frac{3 H_0^2 \Omega_m}{2 c^2} \right) \int_0^{\chi_H} \, \mathrm{d} \chi \, \frac{g_i(\chi) n_j(\chi) F_j\left(\frac{\ell+1/2}{\chi}, \chi\right)}{a(\chi) \, \chi} \, P_\delta \, \left(\frac{\ell + 1/2}{\chi};{\chi} \right) \nonumber \\ 
& \times \mu \left(\frac{\ell+1/2}{\chi}, \chi\right)^2 \left(\frac{1 + \eta\left(\frac{\ell+1/2}{\chi}, \chi\right)}{2}\right) \quad  + \quad {i \leftrightarrow j} \\
P _{i j}^{II} (\ell) &= \int_0^{\chi_H} \, \mathrm{d} \chi \, \frac{n_i(\chi) n_j(\chi) F_i\left(\frac{\ell+1/2}{\chi}, \chi\right) F_j\left(\frac{\ell+1/2}{\chi}, \chi\right)}{\chi^2} P_\delta \left(\frac{\ell + 1/2}{\chi};{\chi} \right) \nonumber \\ & \times \mu\left(\frac{\ell+1/2}{\chi}, \chi\right)^2 , 
\end{align}

while for galaxy-galaxy lensing, the intrinsic alignments contribution is given by:

\begin{align}
P^{gI}_{ij} (\ell) &= \int_0^{\chi_H} \mathrm{d} \chi \, n_j(\chi) n_{Fi}(\chi) F_j \left(\frac{\ell + 1/2}{\chi}, \chi \right) \frac{b_j(\chi)}{\chi^2} \, P_\delta \left(\frac{\ell + 1/2}{\chi};{\chi} \right) \\ &\times  \mu \left(\frac{\ell+1/2}{\chi}, \chi\right)^2. \nonumber
\end{align}

We note that modifications to the linear growth factor $D_+$ and the modified gravity functions $\mu$ and $\eta$ are included for completeness, but their constraining power is limited due to the intrinsic alignment amplitude $A_{\rm IA}$ left free to vary.

\subsection{Priors}

The cosmological parameters in our parameter space are: the cold dark matter density $w_{\mathrm{cdm}}=\Omega_{\mathrm{cdm}}h^2$, the baryon density $w_{\mathrm{b}}= \Omega_{\mathrm{b}}h^2$, the amplitude of the primordial power spectrum $A_s$ (we vary over $\mathrm{ln}10^{10}\mathrm{A_s}$), the value $h$ of the Hubble parameter today divided by 100 km/s/Mpc and the exponent of the primordial power spectrum $n_s$. The parameters $\Omega_{\mathrm{m}},\sigma_8$ and $S_8 = \sigma_8 \sqrt{\Omega_{\mathrm{m}}/0.3}$, shown in the plots in the following Sections, are derived parameters in our analysis. When constraining Horndeski gravity, we vary additionally over either the proportionality coefficients $\hat{\alpha}_B$ and $\hat{\alpha}_M$, for $\alpha_B(\tau) = \hat{\alpha}_B \, \Omega_{\mathrm{DE}}(\tau)$ and $\alpha_M(\tau) = \hat{\alpha}_M \, \Omega_{\mathrm{DE}}(\tau)$, or $\Omega_0$, for the parameterization given by equation~(\ref{eq:planck_parametrization}), where the only free function is $\Omega(\tau) = \Omega_0 \, a(\tau)$. 

The astrophysical nuisance parameters (discussed earlier in Secs.~\ref{sec:nonlinear} and \ref{sec:alignments}) considered in our parameter space are: the intrinsic alignment amplitude $A_{\rm IA}$, the \textsc{HMcode} parameter $c_{\rm min}$ and the effective bias values for the low-$z$ and high-$z$ foreground samples, denoted with $\mathrm{bias}_{z1}$ and $\mathrm{bias}_{z2}$, respectively. We also vary over a cut-off scale for the screening mechanism, as described in Section~\ref{sec:screening}.

We adopt top-hat priors, with ranges specified in Table \ref{tab_prior}, on all cosmological and physical nuisance parameters. We fix $k_{\rm pivot}$, the pivot scale where the scalar spectrum has an amplitude of $A_s$, to 0.05 Mpc$^{-1}$. While massive neutrinos may suppress structure formation in a way that is degenerate with modifications of gravity \citep{Baldi2014, SpurioMancini2018, Reischke2018, Peel2018}, here we fix the sum of the neutrino mass to zero as done in \citet{Hildebrandt2017} and vU18, and adopt their same prior range. All other prior ranges are also set to reproduce the ones used in vU18.

For the Horndeski parameters, when we consider the proportionality to the dark energy density fraction as time dependence for the $\alpha(\tau)$ functions, we vary both $\hat{\alpha}_B$ and $\hat{\alpha}_M$ over the uniform range $[-2, +2]$. This is a symmetric range around the $\Lambda$CDM zero value, chosen such that on one edge, it reaches a point of theoretical singularity for the $\alpha(\tau)$ functions: $\hat{\alpha}_B = 2$ \citep{Noller2018b}. 

When we consider the parameterization for the $\alpha(\tau)$ functions described by equation~(\ref{eq:planck_parametrization}), the only free modified gravity parameter becomes $\Omega_0$, which parameterizes $\Omega(\tau) = \Omega_0 \, a(\tau)$. For this parameter we consider the same prior range $\Omega_0 \in [0,1]$ used in \citet{PlanckCollaboration2016}.

\section{Results}\label{sec:results}
\setlength{\parskip}{0pt}

\begin{figure}
	\includegraphics[width=\columnwidth]{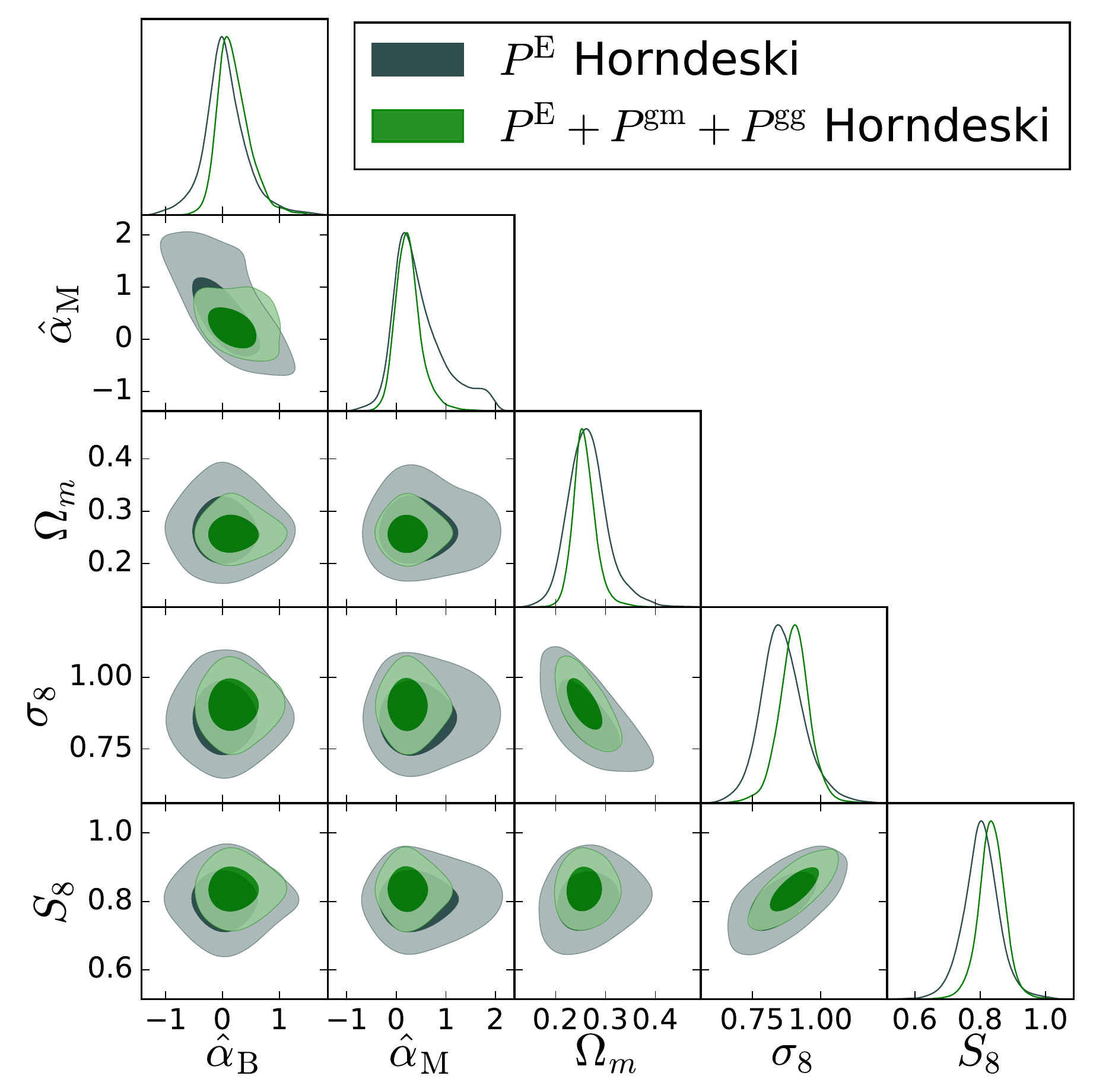}	
	\caption{Comparison of the marginalised 68 and 95 per cent credibility  contours obtained with cosmic shear alone ($P^{\mathrm{E}}$, \textit{grey}) and the three probes combined ($P^{\mathrm{E}}+P^{\mathrm{gm}}+P^{\mathrm{gg}}$, \textit{green}) on $\Omega_{\mathrm{m}}, \sigma_8, S_8$ and the Horndeski parameters $\hat{\alpha}_B$ and $\hat{\alpha}_M$. Note that due to the smoothing applied, the contour in the $\hat{\alpha}_B$-$\hat{\alpha}_M$ plane marginally overlaps with the region forbidden by \textsc{HiClass} stability conditions (cf.~Fig.~\ref{fig:forbidden}).}\label{fig:Plotsuperimposed}
\end{figure}

In this Section we present our cosmological constraints. All of them have been obtained using our new likelihood module for \textsc{MontePython}, which we tested thoroughly against the results obtained by vU18 in a $\Lambda$CDM context. A detailed comparison is described in Appendix \ref{app:comparison}. Here we only comment on the excellent agreement in the parameter $S_8=\sigma_8\sqrt{\Omega_{m}/0.3}$, derived from $\Omega_{\mathrm{m}}$ and $\sigma_8$, i.e. the two parameters whose degenerate combination cosmic shear is most sensitive to \citep{Hildebrandt2017}: for example, the value for $S_8$ that we measure considering only cosmic shear is $0.760^{+0.039}_{-0.038}$, to be compared with $0.761^{+0.040}_{-0.038}$ obtained by vU18. We find excellent agreement with the results of vU18 on all parameters and considering all possible combinations of probes. The fact that our implementation is completely independent from the one in vU18 strengthens both analyses and allows for the use of our likelihood module to obtain constraints on modified gravity, after properly modifying the module so that the mathematical expressions implemented there for the power spectra of the three probes match those for modified gravity introduced in Section~\ref{sec:theory}. 

\begin{figure*}
	\centering
	\includegraphics[scale=0.6]{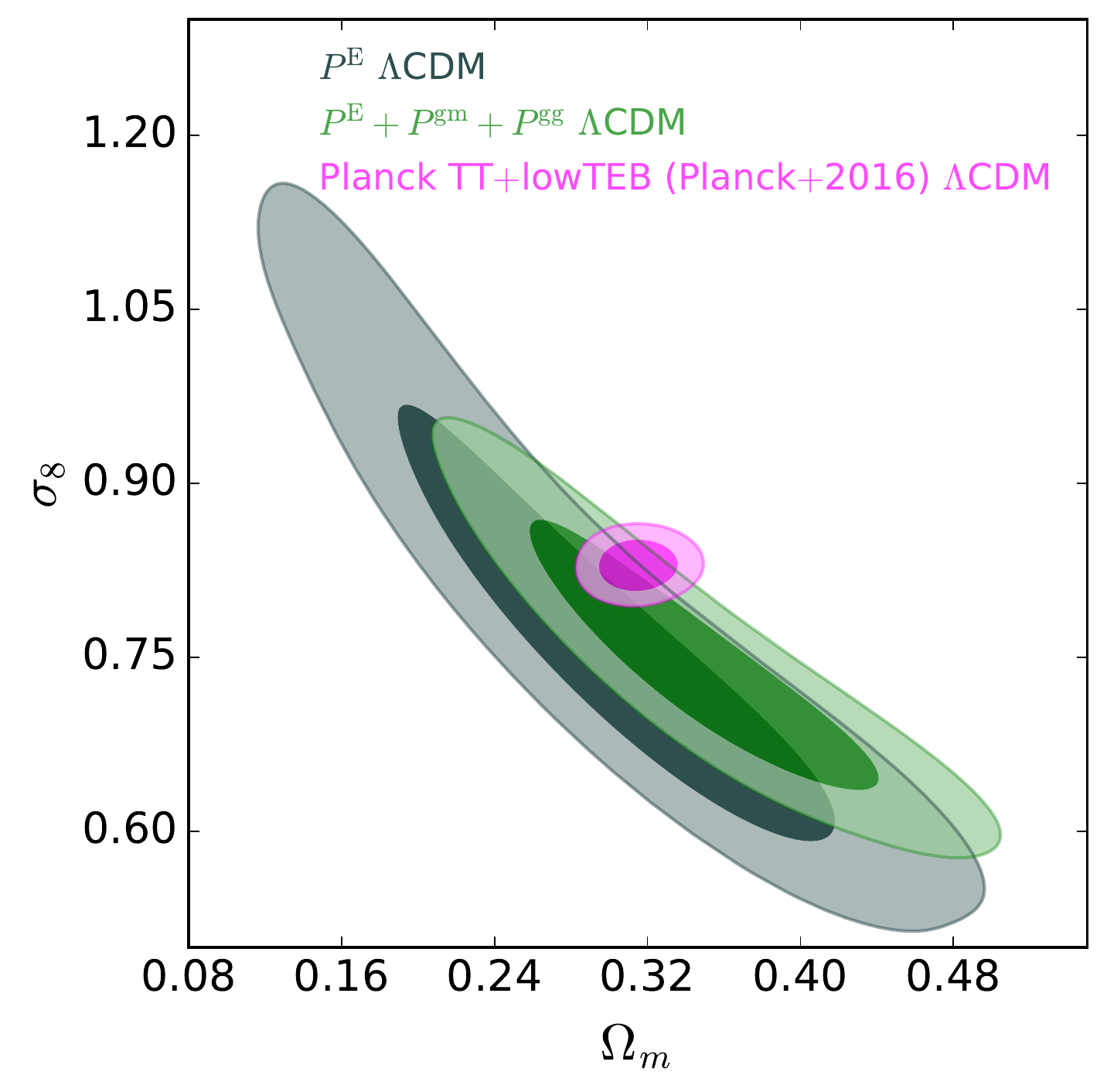}
	\includegraphics[scale=0.6]{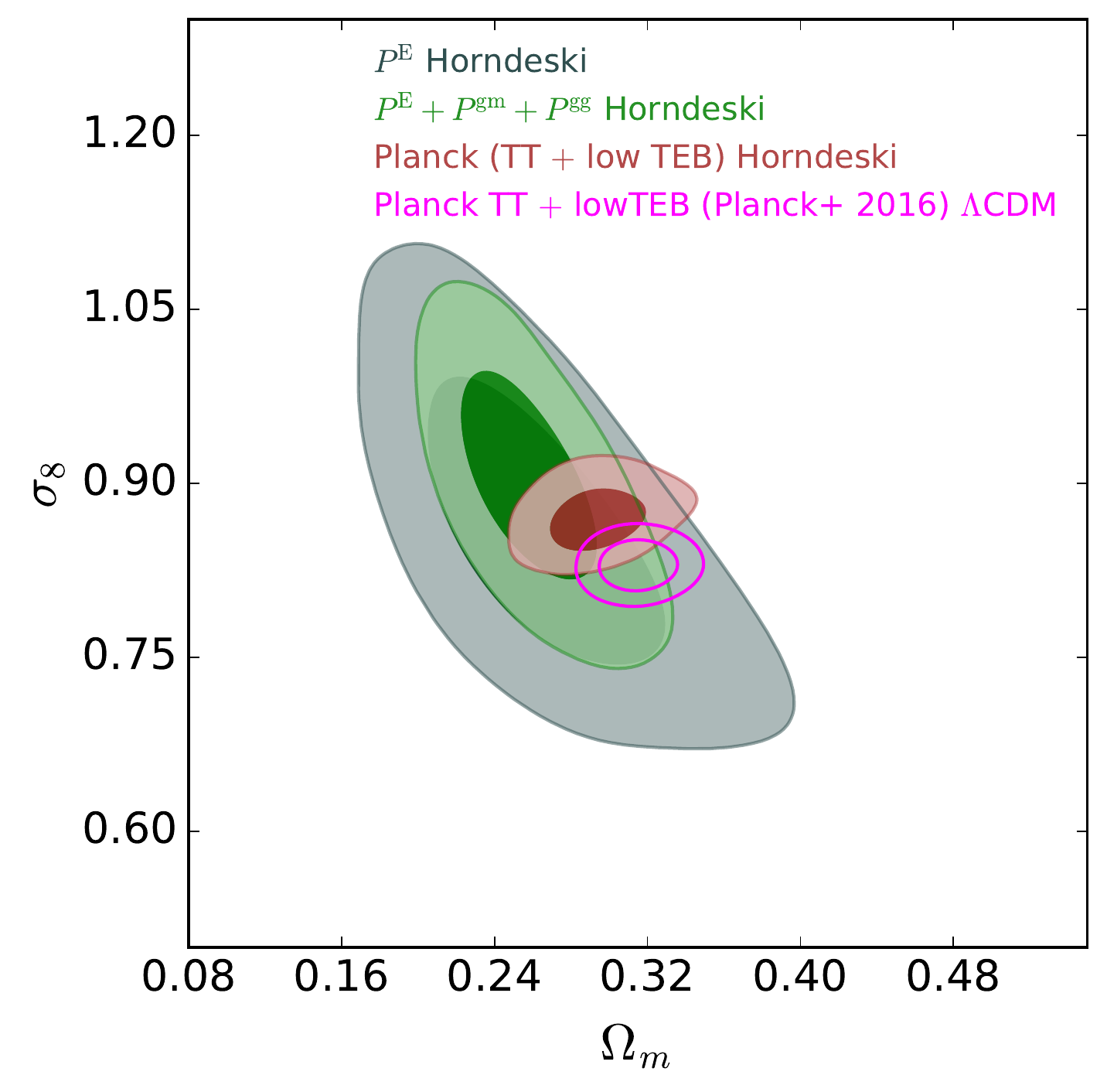}
	\caption{68 and 95 per cent credibility intervals on the $\Omega_{\mathrm{m}}-\sigma_8$ plane obtained with LSS and CMB experiments. In both plots, \textit{grey} contours are obtained with cosmic shear alone, $\textit{green}$ contours with the cross-correlation of cosmic shear - galaxy-galaxy lensing and galaxy clustering. In the \textit{left} panel, LSS and CMB probes are all analysed assuming $\Lambda$CDM model (the \textit{Planck} contours in \textit{magenta} are the ones obtained in \citealt{Planck2016_XIII}). In the \textit{right} panel the LSS constraints are obtained in Horndeski gravity; in \textit{brown} we plot the \textit{Planck} contours, obtained running the \textit{Planck} likelihood in Horndeski gravity. The $\Lambda$CDM contours of \citet{Planck2016_XIII} (the same as in the left panel) are reproduced for comparison, with \textit{magenta} lines.}\label{fig:comparisonLSSPlanck}
\end{figure*}

\subsection{Modified gravity constraints}\label{sec:mg_results}

We investigate the constraining power of our likelihood on dark energy/modified gravity by studying two time parameterizations for the $\alpha(\tau)$ functions, namely the proportionality to the dark energy density fraction described by equation~(\ref{eq:alpha_time_evolution}) and the \texttt{planck\_linear} parameterization introduced in equation~(\ref{eq:planck_parametrization}). Our numerical results for all probe combinations are always compatible with $\Lambda$CDM and are summarised for both parameterizations in Tables ~\ref{tab:ResultsMGproptoOmega} and \ref{tab:ResultsMGLinear}, respectively, showing mean values and 68 per cent credibility intervals. 

We start with the proportionality to $\Omega_{\mathrm{DE}}(\tau)$, and set constraints on the $\hat{\alpha}$ coefficients. Specifically, we consider $\hat{\alpha}_B$ and $\hat{\alpha}_M$ on top of our usual cosmological parameters, while we fix $\alpha_K$ and $\alpha_T$ to zero as explained in Section~\ref{sec:Horndeski}. In Figs.~\ref{fig:bestfitpe},~\ref{fig:bestfitpgm}~and~\ref{fig:bestfitpgg} we overplot the theoretical predictions for the cosmic shear, galaxy - galaxy lensing and galaxy clustering power spectra, respectively, obtained with the best fitting parameters from our combined analysis, in $\Lambda$CDM and in Horndeski gravity; we compare these predictions with the measured power spectra for each probe. We notice how the power spectra produced in Horndeski gravity are characterised by a suppression of the signal towards smaller angular scales, compared to the $\Lambda$CDM spectra. This suggests that our analysis was successful in modelling efficiently the bias factors; if our model had been incorrect (e.g. if the assumption of linear bias had turned out to be insufficient), we would have noticed a steep increase of the power spectra towards higher angular multipoles. As reported in Table~\ref{tab:ResultsMGproptoOmega}, the level of agreement between theoretical predictions and observed data does not vary significantly going from $\Lambda$CDM to Horndeski gravity. While the $\chi^2$ is large in both cosmological scenarios given the relatively high number of degrees of freedom, indicating a bad fit, this is fully explained by a known under-prediction of shot noise contributions to the covariance at large scales (see Section~\ref{sec:cov} for details).

Fig.~\ref{fig:Plotsuperimposed} shows 68 and 95 per cent marginalised contours on the cosmological parameters that our LSS analysis is most sensitive to, again for the parameterization of the $\alpha(\tau)$ functions that sets them proportional to $\Omega_{\mathrm{DE}}(\tau)$. We overplot results obtained considering either cosmic shear alone or all the three probes together. As we can see from the sharp cutoff in the $\hat{\alpha}_B$-$\hat{\alpha}_M$ plane, \textsc{HiClass} applies some stability checks that prevent the Markov Chain Monte Carlo (MCMC) from ending up in regions of parameter space where both of these Horndeski parameters are simultaneously negative (see also Fig.\ref{fig:forbidden}). The investigation of these stability conditions is of the highest priority for future similar analyses \citep[see also][]{Denissenya2018, Noller2018}. We notice that constraints with a multi-probe approach are tighter, as expected; in particular, the value of $S_8$ shrinks from $S_8 = 0.803_{-0.051}^{0.054}$ when measured with cosmic shear alone to $S_8 = 0.835_{-0.039}^{+0.041}$ considering the three probes combined. In both the single and joint analyses the contours for the Horndeski parameters $\hat{\alpha}_B$ and $\hat{\alpha}_B$ are consistent with $\Lambda$CDM values. Specifically, we find $\hat{\alpha}_B = 0.05_{-0.36}^{+0.30} \,$ and $\, \hat{\alpha}_M = 0.46_{-0.61}^{+0.25}$ when considering cosmic shear alone; we retrieve $\hat{\alpha}_B = 0.20_{-0.33}^{+0.20} \,$ and $\, \hat{\alpha}_M = 0.25_{-0.29}^{+0.19}$ in our joint analysis of the three probes. Using cosmic shear alone we find a 68$\%$ marginalised contour in the $\hat{\alpha}_B$-$\hat{\alpha}_M$ plane that covers 11 per cent of the parameter space that is allowed to be explored by the stability conditions applied by \textsc{HiClass}. The same contour obtained with the joint three-probe analysis shrinks down to 6 per cent. Similarly to vU18, we confirm that the increased constraining power with the multi-probe approach comes from the tighter constraints achievable on nuisance parameters. The inclusion of the cross-correlation with the other probes tightens the constraints in particular on the intrinsic alignment amplitude, mostly constrained by the inclusion of galaxy-galaxy lensing in the analysis, which shrinks the error on $A_{IA}$ from $0.44^{+1.16}_{-0.68}$ for cosmic shear alone, to $1.37^{+0.48}_{-0.44}$ with the multi-probe approach. This decrease in the error for $A_{\mathrm{IA}}$ is very similar to the one found by vU18 in their $\Lambda$CDM analysis; in both $\Lambda$CDM and Horndeski scenarios, the cross-correlation of the three probes shrinks the error on the intrinsic alignment amplitude to approximately half its cosmic shear-only value. At the same time, the best fit value for $A_{\mathrm{IA}}$ doubles in the multi-probe approach with respect to the cosmic shear analysis, making the measurement four times more significant. Our combined fit in Horndeski gravity constrains very well the bias parameters for the two foreground samples, similarly to what found in vU18; interestingly, however, our errors in Horndeski gravity are half of those reported by vU18 in $\Lambda$CDM. Constraints on $\sigma_8$ and $\Omega_{\mathrm{m}}$ become tighter in the combined analysis, and consequently the same happens for $S_8$.

The second parameterization for the $\alpha(\tau)$ functions that we consider is the one called \texttt{planck\_linear} in \textsc{HiClass}, which we already introduced in Section~\ref{sec:Horndeski}. In this parameterization the four $\alpha$ functions can be specified through a single function of time $\Omega\left(\tau\right)$ affecting three of the $\alpha$ functions (see equation~\ref{eq:planck_parametrization}). In this case we do not observe tighter constraints on the modified gravity parameter $\Omega_0$ from the inclusion of the multi-probe cross-correlation, while we do find stronger constraints on $\Omega_{\mathrm{m}}, \sigma_8$ and $S_8$, similarly to the case of Horndeski functions proportional to $\Omega_{\mathrm{DE}}(\tau)$. For both parameterizations considered for the $\alpha(\tau)$ functions we find that the screening scale $k_s$ is largely unconstrained.

\subsection{Comparison and combination with \textit{Planck}}\label{sec:comparison_Planck}

\begin{figure*}
	\begin{multicols}{2}
		\includegraphics[width=\linewidth]{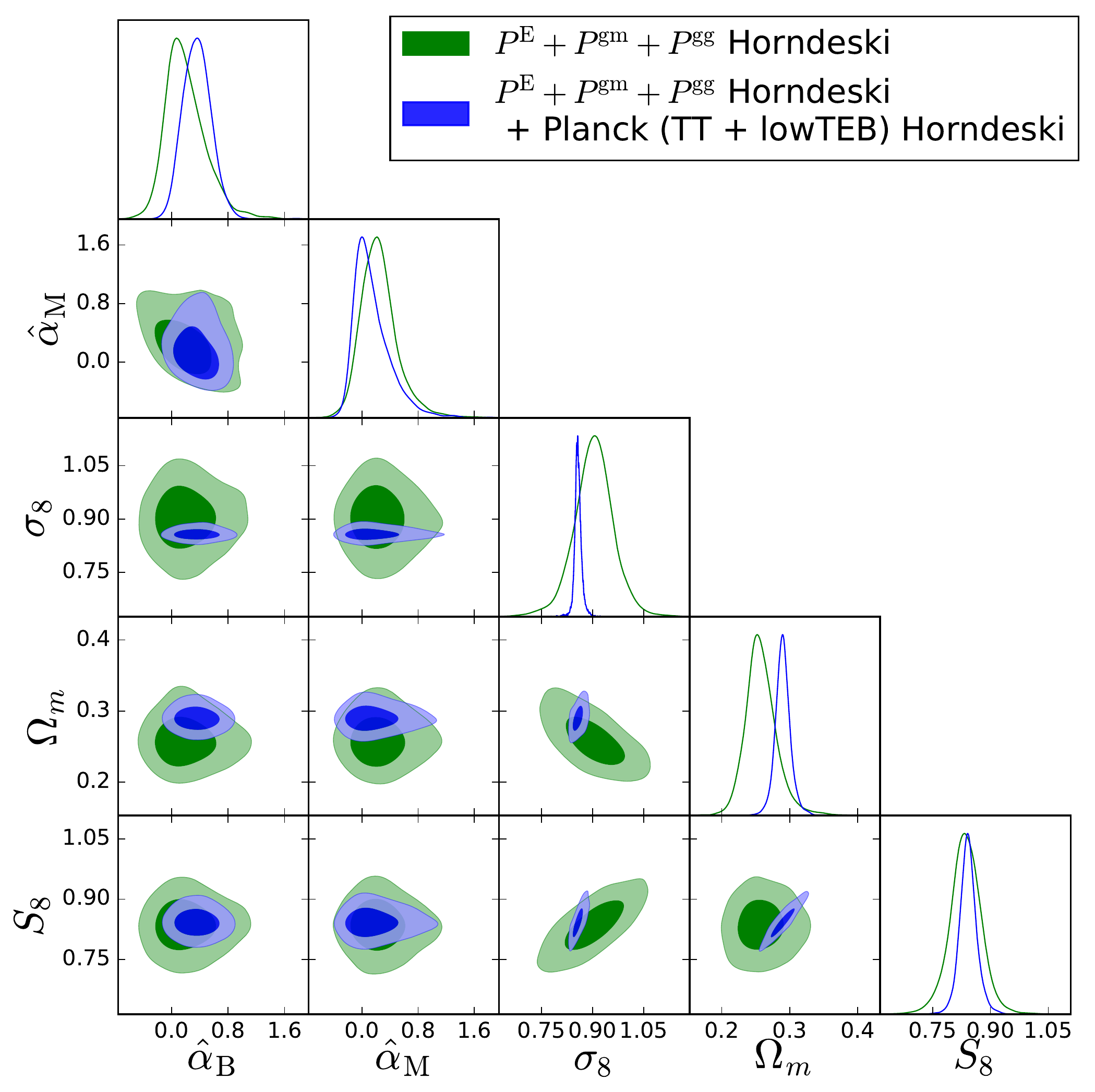}\par 
		\includegraphics[width=\linewidth]{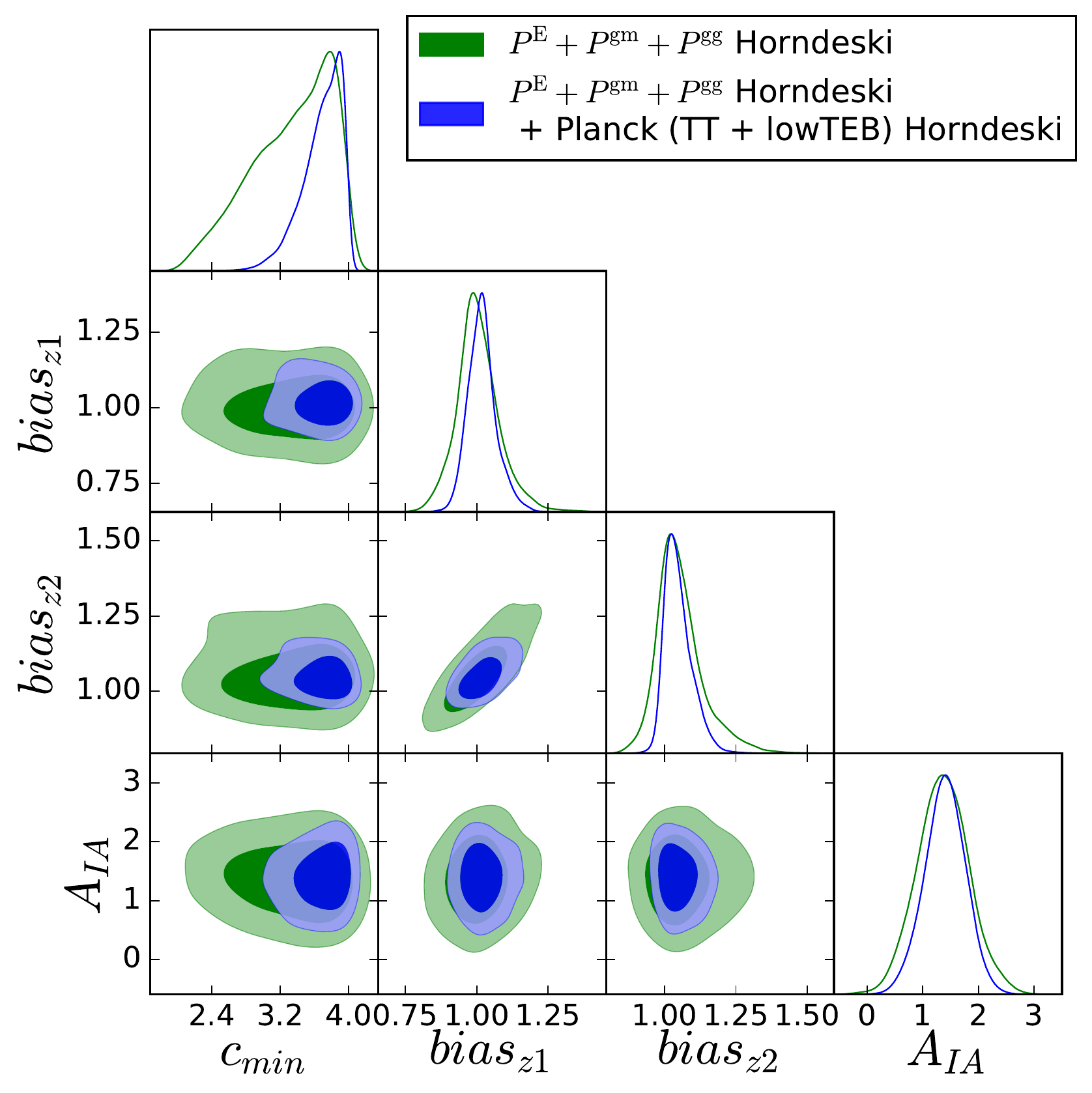}\par
	\end{multicols}
	\caption{Increase in sensitivity to key cosmological parameters (\textit{left} panel) and astrophysical nuisance parameters (\textit{right} panel) that results from running the MCMC of the joint cosmic shear -galaxy-galaxy lensing - galaxy clustering (\textit{green}) in combination with CMB \textit{Planck} likelihood (\textit{blue}). The nuisance parameters (discussed in Secs.~\ref{sec:nonlinear} and \ref{sec:alignments}) are the \textsc{HMcode} parameter $c_{\mathrm{min}}$, the effective bias values $\mathrm{bias}_{z1}$ and $\mathrm{bias}_{z2}$ for the low-$z$ and high-$z$ foreground samples, respectively, and the intrinsic alignment amplitude $A_{\mathrm{IA}}$.}\label{fig:pepgmpggPlanck}
\end{figure*}

We investigated the relevance of our analysis in the context of the tension between cosmological parameters estimated with weak gravitational lensing analyses and CMB measurements. Fiducial analyses of the CFHTLenS \citep{Heymans2013, Joudaki2017} and KiDS \citep{Hildebrandt2017, Kohlinger2017} collaborations have reported values for $\Omega_{\mathrm{m}}$ and $\sigma_8$ that are in mild tension with those measured by \citet{Planck2016_XIII}. Here and throughout our analysis we will refer to the 2015 \textit{Planck} results because in that case the likelihood modules, to be run with \textsc{MontePython}, are publicly available, thus allowing for re-running of the MCMC. We notice, however, that the latest \textit{Planck} results \citep{Planck2018} do not differ significantly from the earlier release.

For comparison with \textit{Planck}, we choose the combination of two likelihoods from the \textit{Planck} 2015 release: the {\tt Plik\_lite} likelihood for the temperature-only power spectrum in the range $\ell = 30-2508$, and the {\tt lowTEB} likelihood for temperature and LFI polarization information in the range $\ell=2-29$. We choose this likelihood as it is the same used in vU18. For the high-$\ell$ temperature likelihood, we run the \texttt{Plik lite} likelihood, which is pre-marginalised over all nuisance parameters except for the \textit{Planck} absolute calibration. \citet{Noller2018b} demonstrated that running the full high-$\ell$ temperature likelihood or its pre-marginalised version produces equivalent constraints on Horndeski parameters. However, the \texttt{Plik\_lite} version produces a remarkable speed up in the analysis, as the number of nuisance parameters for the \textit{Planck} part is reduced to only one. We have in common with the analysis of \citet{Noller2018b} a pure $\Lambda$CDM background which, as the authors of that paper suggest, is one of the reasons behind the excellent agreement between the \texttt{Plik\_lite} and full \textit{Planck} likelihood, since the \texttt{Plik\_lite} likelihood is pre-marginalised assuming a $\Lambda$CDM cosmology. However, as also reported in \citet{Noller2018b}, we remark that extensions of our analysis considering background cosmologies different from $\Lambda$CDM should arguably run the \textit{Planck} likelihood in its full version.    

\begin{figure}
	\centering
	\includegraphics[width=\columnwidth]{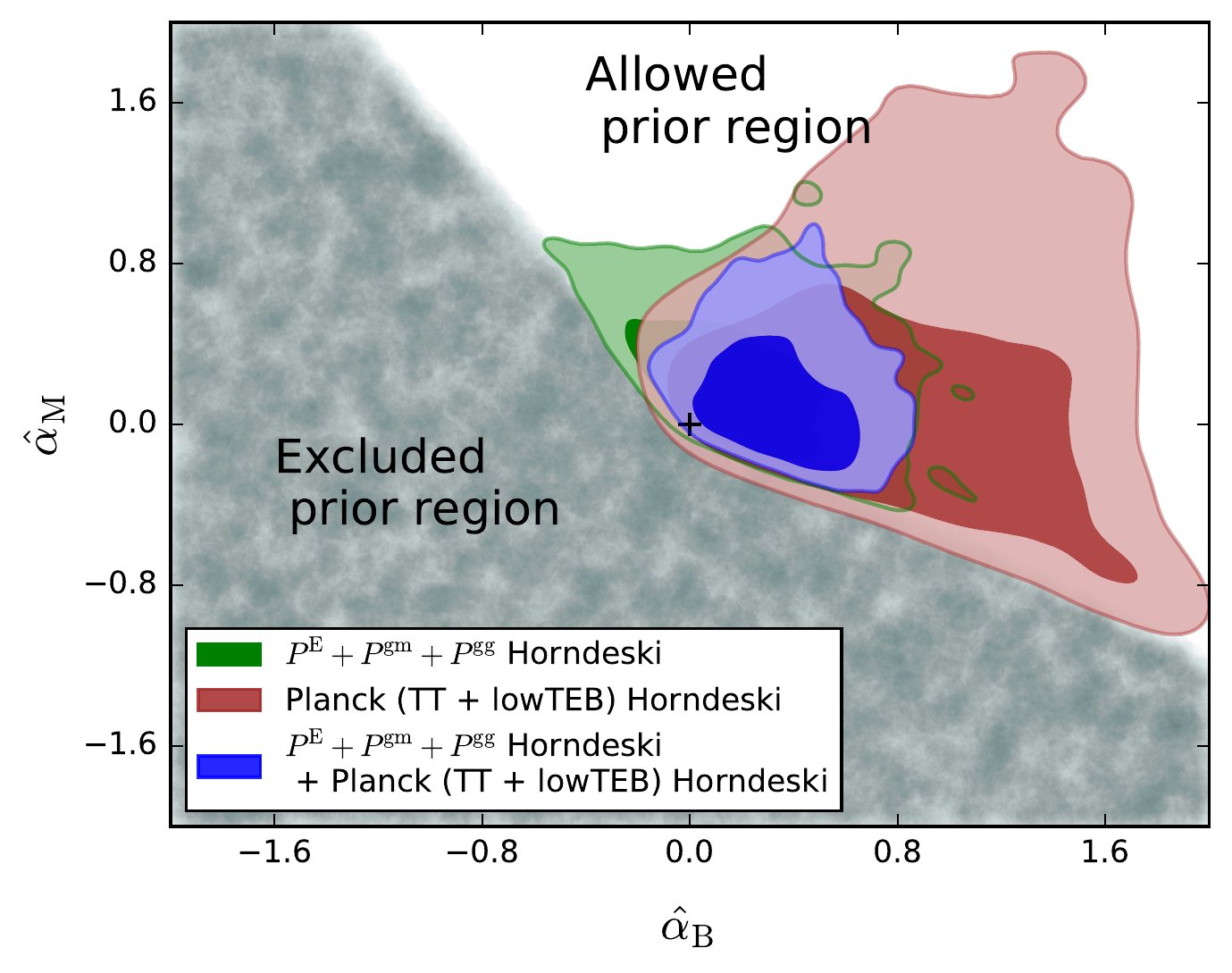}
	\caption{Region of parameter space (in \textit{white}) in the plane $\hat{\alpha}_B - \hat{\alpha}_M$ that is allowed to be explored by the MCMC, and 68 and 95 per cent marginalised contours in the same plane, for different experiments. We plot in \textit{green} the LSS constraints from our joint cosmic shear, galaxy-galaxy lensing and galaxy clustering; in \textit{brown}, we represent the same contours for the \textit{Planck} CMB likelihood. The \textit{blue} contours are those obtained running together the LSS and CMB likelihoods. The stability conditions applied by \textsc{HiClass} do not allow the MCMC to end up in the \textit{grey} region (the marginal overlap of the contours with the excluded region is only due to smoothing). The fact that contours are centered slightly away from the cross-marked $\Lambda$CDM point $\{0,0\}$ (while still being consistent with the $\Lambda$CDM prediction) is due to the skewness of the marginal distributions of these parameters, a consequence of \textsc{HiClass} stability conditions.}\label{fig:forbidden}
\end{figure}
We compare the results of our LSS analysis, which considers cross-correlations of our three probes, with \textit{Planck} results. This comparison can be summarised as in Fig.~\ref{fig:comparisonLSSPlanck}, where the two subplots show results for different choices of cosmological model in this comparison. Inference results are presented for $\Omega_{\mathrm{m}}$ and $\sigma_8$, the two cosmological parameters that are best constrained by cosmic shear. In our analysis these two parameters are derived from the cold dark matter density $w_{\mathrm{cdm}} = \Omega_{\mathrm{cdm}} h^2$ and the rescaled amplitude of the primordial power spectrum $\ln(10^{10}A_{\rm s})$. In the left panel, we show the results of running our LSS probes in $\Lambda$CDM and overplotting the constraints in the $\Omega_{\mathrm{m}}-\sigma_8$ plane with those obtained by \citet{Planck2016_XIII} also in $\Lambda$CDM. This plot essentially reproduces Fig.~7 in vU18, but the LSS contours are obtained with our own, independently developed likelihood: we recognise the familiar mild `tension' between LSS and CMB measurements. In the right panel of Fig.~\ref{fig:comparisonLSSPlanck} we analyse the LSS probes in a Horndeski scenario, and also re-run the \textit{Planck} likelihood in Horndeski gravity. The agreement between LSS and CMB contours is excellent. For comparison, in the same $\Omega_{\mathrm{m}} - \sigma_8$ plane we overplot the same $\Lambda$CDM contours for \textit{Planck} presented in the left subplot, corresponding to the results published in \citet{Planck2016_XIII}. We notice that there is already a mitigation of the tension between LSS and CMB results when one considers the LSS constraints in Horndeski gravity, and the CMB ones in $\Lambda$CDM. The alleviation of the tension, compared to the $\Lambda$CDM results, originates from the combined effect of the larger Horndeski parameter space, which widens the LSS contours in the $\Omega_{\mathrm{m}}-\sigma_8$ plane so that they better match with the $\Lambda$CDM contours from \textit{Planck}, as well as a shift in the best fit value for the parameters.

To measure the agreement between the $S_8$ parameter retrieved by our Horndeski and $\Lambda$CDM gravity analyses, in comparison to the \textit{Planck} values, we calculate the shift $\Delta S_8$ between this parameter measured by the CMB and LSS probes. Considering cosmic shear alone, we report a value $\Delta S_8 = 0.048_{-0.056}^{+0.059}$ in Horndeski, compared to $\Delta S_8 = 0.091_{-0.045}^{+0.046}$ in $\Lambda$CDM. In the joint three-probe analysis, we find $\Delta S_8 = 0.016_{-0.046}^{+0.048}$ in Horndeski gravity and $\Delta S_8 = 0.059_{-0.039}^{+0.040}$ in $\Lambda$CDM.

As a last step in the comparison of our LSS analysis with the \textit{Planck} CMB results, we combine the LSS and CMB likelihoods. When doing this, we initially extended the \textit{Planck} chains to the inclusion of CMB lensing; while this could in principle induce correlations between \textit{Planck} and KiDS, we found that CMB lensing had no impact on the results presented; therefore, we decided to report results obtained without the addition of CMB lensing, as they do not differ significantly from the ones obtained with CMB lensing. Crucially, we remark here that we are allowed to combine the LSS and CMB likelihoods because, contrary to what happened when analysing LSS and CMB probes in $\Lambda$CDM (as in \citealt{Hildebrandt2017} and vU18), our Horndeski analysis produces constraints on the cosmological parameters, for LSS and CMB, that are not in tension between them, as shown in Fig.~\ref{fig:comparisonLSSPlanck}. 

The two panels of Fig.~\ref{fig:pepgmpggPlanck} show the increase in sensitivity of the combined LSS+CMB analysis to key cosmological parameters (left panel) including the Horndeski parameters $\hat{\alpha}_{B}$ and $\hat{\alpha}_{M}$, and astrophysical nuisance parameters (right panel) such as the bias coefficients, the intrinsic alignment amplitude and the \textsc{HMcode} parameter $c_{\mathrm{min}}$, resulting from running our LSS chains in combination with the \textit{Planck} CMB ones. We find that, as expected, combining LSS and CMB datasets reduces uncertainties on all cosmological parameters. In particular, \textit{Planck} measurements allow for better constraints on parameters that are weakly constrained by LSS probes, such as $n_s$ and $h$ (see Table \ref{tab:ResultsMGproptoOmega}). The Horndeski parameters $\hat{\alpha}_B$ and $\hat{\alpha}_M$ are also better constrained in the combined analysis, although we notice that most of the constraining power on these two parameters comes from the LSS probes. This can be seen better in Fig.~\ref{fig:forbidden}, where we plot the region of parameter space in the $\hat{\alpha}_B-\hat{\alpha}_M$ plane that is allowed to be explored by the stability conditions applied by \textsc{HiClass} \citep[which ensure that the choice of the Horndeski parameters corresponds to a theoretical model that is safe from ghost, gradient and tachyon instabilities; for details, see][]{Zumalacarregui2016}, as well as the marginalised 68 and 95 per cent contours that can be obtained on these parameters using either our joint LSS analysis, re-running the \textit{Planck} likelihood in Horndeski gravity or running together the \textit{Planck} and LSS likelihoods. The combination of LSS and CMB likelihoods increases the sensitivity to the Horndeski parameters, with the 68 per cent contour covering 5 per cent of the allowed parameter space in the $\hat{\alpha}_B$ - $\hat{\alpha}_M$ plane, although the contours obtained with the combined LSS+CMB analysis are comparable in size with the ones obtained from the LSS probes alone. We remark that the contours appear to be centered away from the $\Lambda$CDM point $\{0,0\}$ (while still being consistent with this value) due to the skewness in the marginal distributions of the parameters; this is due to stability conditions applied by \textsc{HiClass}, which artificially cut out a part of parameter space. A plot comparing the contours on all cosmological and nuisance parameters obtained with our LSS analysis and those produced running the LSS and \textit{Planck} likelihood together can be found in Appendix \ref{app:full_posterior}. The numerical values for the mean and 68 per cent credibility intervals for all cosmological and nuisance parameters in the combined LSS+CMB analysis are reported in Table \ref{tab:ResultsMGproptoOmega}, to be compared with the LSS-only and \textit{Planck}-only results, also reported in Table \ref{tab:ResultsMGproptoOmega}. The exact number of degrees of freedom for the \textit{Planck} likelihood is not readily available in the literature \citep[see e.g.][]{Planck2016_XIII} and its non-trivial calculation is beyond the scope of this work. We reproduced the value of $\chi^2$ obtained by the official \textit{Planck} analysis in $\Lambda$CDM, which we found to be very similar to the one we obtained in Horndeski gravity. In the combined large-scale structure and CMB analysis we notice that the large-scale structure probes add a negligible contribution to the total number of degrees of freedom compared to \textit{Planck}; since we find that the $\chi^2$ for the combined large-scale structure and CMB  analysis increases by $\sim 1\%$ from the CMB-only value, we conclude that there is no indication that the goodness of fit degrades in the combined analysis with respect to considering \textit{Planck} alone. 

\section{Discussion and conclusions}\label{sec:conclusions}

In this paper we analysed data from the KiDS and GAMA surveys, in a joint framework for tomographic cosmic shear, galaxy-galaxy lensing and angular clustering power spectra. Our numerical implementation reproduces the $\Lambda$CDM results of vU18 and extends their analysis to a Horndeski scenario: we set constraints on both standard cosmological parameters as well as functions that fully describe the evolution of linear perturbations in Horndeski gravity. We considered two different time parameterizations for these functions: one where they are proportional to the dark energy density fraction and another one where there is effectively only one free function which we set proportional to the scale factor; the latter represents a subset of models within the Horndeski class. 

For all the combinations of probes considered, we found that our constraints are compatible with $\Lambda$CDM values: specifically, all Horndeski parameters are compatible with zero (their GR value) within their 68 per cent marginalised errors.~We notice that while the inclusion of cross-correlations of the three probes tightens the constraints on both cosmological and nuisance parameters, as expected our constraints are rather large \textit{per se}, for multiple reasons. Our parameter space is very large, because we do not fix any background values for the standard cosmological parameters; we vary over them as well. Additionally, while the KiDS data that we used represent state-of-the-art imaging data that are currently available from a cosmic shear survey, the survey volume considered is still relatively small; we do expect increased constraining power from larger data releases. We remark that, while the constraining power of the datasets considered in our analysis is not sufficient to produce very strong constraints on the values of the Horndeski $\alpha(\tau)$ functions (or more precisely, on the parameters that describe one of the possible time parameterizations for these functions), there are some important conclusions that we can draw from the constraints we found.

In particular, assuming the commonly studied relation of proportionality with $\Omega_{\mathrm{DE}}$ for the $\alpha(\tau)$ functions, we found that the values of both proportionality coefficients $\hat{\alpha}_B$ and $\hat{\alpha}_M$ are close to zero and smaller than unity. Following the line of reasoning presented in \cite{Bellini2016}, if one assumes the time parameterization of proportionality to $\Omega_{\mathrm{DE}}(\tau)$, one would expect the coefficients $\hat{\alpha}$ to be $\mathcal{O}(1)$, because in this parameterization the $\alpha(\tau)$ functions are driven by the same functions of the scalar field as the energy density and its derivatives. The fact that we found values for the proportionality coefficients $\hat{\alpha}$ significantly smaller than unity can, in this sense, be regarded as a substantial observational constraint on dark energy/modified gravity derived from this analysis. 

We compared and combined our LSS constraints on Horndeski gravity with the \textit{Planck} analysis of the CMB. Running our LSS and the \textit{Planck} CMB likelihood separately, we found that the enlarged Horndeski parameter space for the LSS probes helps reduce the tension in cosmological parameters with the CMB measurements obtained in $\Lambda$CDM; the agreement becomes excellent if we compare our LSS Horndeski results with the \textit{Planck} chains re-run in Horndeski gravity. Since separately these LSS and \textit{Planck} constraints in Horndeski gravity are not in tension, we are allowed to combine the LSS and CMB likelihoods. The constraints obtained by this combined analysis are tighter as expected, although we notice that most of the constraining power on the Horndeski parameters comes from the LSS likelihood; the Planck likelihood helps constrain much better other standard cosmological parameters, such as $n_s$ and $h$, that are weakly constrained by the LSS probes. When combining LSS and CMB likelihoods, we keep the assumption of flatness that we also imposed on our LSS-only analysis. While flatness is predicted by inflation, future analyses should allow for the possibility of non-vanishing curvature when combining LSS and \textit{Planck} likelihoods.    

For the non-linear matter power spectrum we followed the official KiDS prescription, using the \citet{Mead2015} correction implemented in \textsc{HMcode}, which in turn is based on \textsc{halofit}. A bespoke, accurate prescription for the non-linear matter power spectrum in the full model space covered by Horndeski gravity is not yet available. However, note that we are exploring constraints around the $\Lambda$CDM values, where \textsc{halofit} is fully calibrated, and that we apply the non-linear correction to the linear power spectrum rigorously calculated by \textsc{HiClass} in a modified gravity scenario. We argue that the physical mechanism implemented in \textsc{halofit} produces a reasonable degree of predictive power on the extended cosmological models considered, appropriate for the currently still fairly low constraining power of the data. This statement is corroborated by the analysis in Appendix \ref{app:nonlinearities}, which shows quantitatively the fairly low sensitivity of our data to the modelling uncertainty of the non-linear matter power spectrum in Horndeski gravity. Similarly, the screening mechanism has been implemented in a phenomenological way due to a lack of an evolved screening mechanism in the linear formalism of the $\alpha(\tau)$ functions. We remark that future work in these directions is needed (see e.g.~\citealt{Cataneo2019}) to provide tighter constraints with either larger data releases of the current survey generation or future data from Stage IV surveys such as \textit{Euclid}~\citep{Laureijs2011}\footnote{\url{https://www.euclid-ec.org/}} and the Large Synoptic Survey Telescope \citep{LSSTScienceCollaboration2009}\footnote{\url{https://www.lsst.org/}}. 

One point to highlight is the difficult interpretation of parameter space as far as the Horndeski functions are concerned.~Assuming proportionality to $\Omega_{\mathrm{DE}}$, in our analysis these parameters were let free to vary over negative values as well. However, we found a very distinctive cutoff in the $\hat{\alpha}_{\mathrm{B}}$-$\hat{\alpha}_{\mathrm{M}}$ plane, which prevents the chains to end up in regions of parameter space where both proportionality coefficients are significantly negative. This is due to some stability checks (in particular those concerning the positivity of the speed of sound) which reject points ending up in those regions of parameter space. Those regions are, however, interesting because they may refer to models of relevance, such as some $f(R)$ models for which the relation $\alpha_B = -\alpha_M$ holds. While we are aware that some of these stability conditions may in some cases be harmlessly bypassed, because they reject points in parameter space even when they produce a negative speed of sound \textit{only} at very early times, the exploration of these stability conditions is an area of active research. The recent analysis of \citet{Denissenya2018}, for example \citep[but see also a discussion on this in][]{Kreisch2017}, has investigated the relation between parameterization and stability in Horndeski gravity, showing that the relation is not trivial. In particular, they have shown how stability evolves with redshift, picking out different regions of parameter space that can have complex structure. These end up forming disconnected 'islands' in parameter space, which are significantly sensitive to the time evolution assumed for the $\alpha(\tau)$ functions.

The very recent analysis of \citet{Noller2018} suggests that radiative stability places strong constraints on the Horndeski parameter space; they used this condition to obtain tight constraints from the analysis of data from the \textit{Planck}, SDSS/BOSS and 6dF surveys.~\citet{Noller2018b} also presented cosmological constraints on Horndeski gravity using CMB, redshift space distortions, matter power spectrum and BAO measurements from the \textit{Planck}, SDSS/BOSS and 6dF surveys. While our results are obtained considering different cosmological probes with respect to theirs, interestingly our constraints compare well with those presented in their analysis; this is true in particular in the $\hat{\alpha}_B-\hat{\alpha}_M$ plane, when considering the time dependence for the Horndeski parameters that makes them proportional to the dark energy density fraction.~We remark that the investigation of the theoretical stability conditions for the $\alpha(\tau)$ functions seems a high priority issue to be addressed in future extensions of our analysis.

To conclude, we stress that, in view of future work, the importance of our KiDS+GAMA analysis lies in two main aspects. On the one hand, we have developed a first likelihood module for Horndeski gravity that can analyse all possible cross-combinations of cosmic shear, galaxy-galaxy lensing and galaxy clustering power spectra data, as can be produced by current and future surveys. On the other hand, our likelihood module can also be run in a standard $\Lambda$CDM scenario and in this regard our results are in excellent agreement with those from the fully independent analysis pipeline of vU18, representing a benchmark for our implementation as well as a further cross-check of their fiducial KiDS+GAMA analysis. In the future it will be interesting to carry out a similar analysis with larger data releases from the KiDS survey and/or with data from Stage IV surveys such as \textit{Euclid}; we expect that the work presented in this paper and its publicly available implementation will represent a useful tool for these or similar analyses.

\section*{Acknowledgements}

ASM acknowledges financial support from the graduate college \textit{Astrophysics of cosmological probes of gravity} by Landesgraduiertenakademie Baden-W\"urttemberg. ASM and BJ also acknowledge support by the UCL Cosmoparticle Initiative. We thank the German DFG Excellence Initiative for funding a mobility program, through the grant \textit{Gravity on the largest scales and the cosmic large-scale structure}, within which this work originated. 
FK acknowledges support from the World Premier International Research Center Initiative (WPI), MEXT, Japan and from JSPS KAKENHI Grant Number JP17H06599.
Most of the numerical results of this work have been derived with software run on the Hypatia cluster at University College London: we are grateful to Hiranya Peiris for making this possible and to Edd Edmondson for the technical support with it. We thank Carlos Garc\'{i}a Garc\'{i}a, Emilio Bellini and Miguel Zumalac\'{a}rregui for their help with \textsc{MontePython} and Pedro Ferreira, Steven Gratton, Shahab Joudaki, Gra\c{c}a Rocha and Patricio Vielva for useful discussions. 

Based on data products from observations made with ESO Telescopes at the La Silla Paranal Observatory under programme IDs 177.A-3016, 177.A-3017 and 177.A-3018. We use cosmic shear measurements from the Kilo-Degree Survey \citep{Kuijken2015, Hildebrandt2017, FenechConti2017}, hereafter referred to as KiDS. The KiDS data are processed by THELI \citep{Erben2013} and Astro-WISE \citep{Begeman2013, deJong2015}. Shears are measured using \textsc{lensfit} \citep{Miller2013}, and photometric redshifts are obtained from PSF-matched photometry and calibrated using external overlapping spectroscopic surveys \citep{Hildebrandt2017}. GAMA is a joint European-Australasian project based around a spectroscopic campaign using the Anglo-Australian Telescope. The GAMA input catalogue is based on data taken from the Sloan Digital Sky Survey and the UKIRT Infrared Deep Sky Survey. Complementary imaging of the GAMA regions is being obtained by a number of independent survey programs including GALEX
MIS, VST KiDS, VISTA VIKING, WISE, Herschel-ATLAS, GMRT, and ASKAP providing UV to radio coverage. GAMA is funded by the STFC (UK), the ARC (Australia), the AAO, and the participating institutions. The
GAMA website is \url{http://www.gama-survey.org/}.

\bibliographystyle{mnras}
\bibliography{references} 

\appendix

\section{Derivation of modified gravity power spectrum for cosmic shear}\label{app:cosmicshear_mg_power}

Here we present a detailed derivation of equation~(\ref{eq:pe}), i.e. the cosmic shear power spectrum for a generic modified gravity theory parameterised by the functions $\mu$ and $\eta$ introduced in Eqs.~\ref{eq:mu} and \ref{eq:eta}. Expressions for galaxy-galaxy lensing and galaxy clustering (corresponding to Eqs.~\ref{eq:pgm} and \ref{eq:pgg}) can be derived following a similar procedure. 

Cosmic shear power spectra can be most elegantly derived from expressions for the lensing potential power spectra, since the shear is linearly defined in terms of the lensing potential, through its second derivatives. Moreover, to avoid complications arising from the spin-2 tensorial nature of the shear field, we can compute power spectra for the convergence, which is a scalar field similarly related in a linear way to the lensing potential. Convergence and shear power spectra are equal in the flat sky approximation and differ only as a result of a different multipole-dependent prefactor \citep{Kilbinger2017}, which for $\ell > 45$ represents a difference of less than $0.1\%$ between the two power spectra; given our multipole range, we can safely ignore this difference.

As remarked in Section~\ref{sec:theory}, a generic modified gravity scenario is characterised by Bardeen potentials $\Phi$ and $\Psi$ in equation~(\ref{def:pot}), which can differ from each other. Thus, the lensing potential $\phi_i$ in tomographic bin $i$ can be defined as 
\begin{align}
\phi_i = \frac{1}{c^2} \int \mathrm{d} \chi \left( \Phi+\Psi \right) g_i (\chi), 
\end{align} 
where $g_i$ is the geometric weight factor for bin $i$ introduced in equation~(\ref{eq:g_i}), and the integration is carried out along the line of sight in Born approximation. The correlation of the lensing potential in tomographic bins $i$ and $j$ is thus
\begin{align}\label{eq:correlation_potential}
\left\langle \phi_i \phi_j \right\rangle (\ell) = \frac{1}{c^4} \int \frac{\mathrm{d}\chi}{\chi^2}g_i(\chi)g_j(\chi)P_{\Phi+\Psi}\left( \frac{\ell+1/2}{\chi}; \chi \right), 
\end{align}
where we made use of the extended Limber approximation, and $P_{\Phi+\Psi}$ is the power spectrum of the sum of the Bardeen potentials. Since the convergence $\kappa$ is related to the Laplacian of the lensing potential:
\begin{align}\label{eq:convergence}
\Delta \phi = 2 \kappa\;,
\end{align}
rewriting equation~(\ref{eq:correlation_potential}) in terms of the convergence gives us 
\begin{align}
\left\langle \kappa_i \kappa_j \right\rangle (\ell) = \frac{1}{4 c^4} \int \frac{\mathrm{d}\chi}{\chi^2}g_i(\chi)g_j(\chi) \frac{P_{\Phi+\Psi}}{k^4}
\end{align}
where the $k^{-4}$ factor comes from equation~(\ref{eq:convergence}) expressed in Fourier space. Note that this is a two dimensional Fourier vector on the flat sky. It can be augmented by a third dimension in the thin lens approximation \citep{Bartelmann2001}.
Using now the (modified) Poisson equation~(\ref{eq:mu}), to link the gravitational potential to the overdensity field, and recalling that the two Bardeen potentials $\Phi$ and $\Psi$ are related through $\eta$ (equation~\ref{eq:eta}), we find that
\begin{align}
\frac{P_{\Phi+\Psi}\left( \frac{\ell+1/2}{\chi} \right)}{k^4} = \left( \frac{3}{2} \frac{\Omega_{\rm m} H_0^2}{c^2} \right)^2 P_{\delta} \left( \frac{\ell+1/2}{\chi} \right) \left[ 1 + 2\eta + \eta^2 \right] \mu^2, 
\end{align}
with $P_\delta$ the matter power spectrum.
Thus, 
\begin{align}
\left\langle \kappa_i \kappa_j \right\rangle (\ell) = &\left( \frac{3\Omega_{\rm m} H_0^2}{2c^2} \right)^2 \frac{1}{4} \int \frac{\mathrm{d}\chi}{\chi^2} \frac{g_i(\chi)}{a} \frac{g_j(\chi)}{a} \mu^2 \left[ 1 + 2 \eta +\eta^2 \right] P_\delta \left( \frac{\ell+1/2}{\chi} \right),
\end{align}
which leads to equation~(\ref{eq:pe}). 

\section{Full posterior distribution}\label{app:full_posterior}
\setlength{\parskip}{0pt}

\begin{figure*}
	\makebox[\textwidth][c]{\includegraphics[scale=0.37]{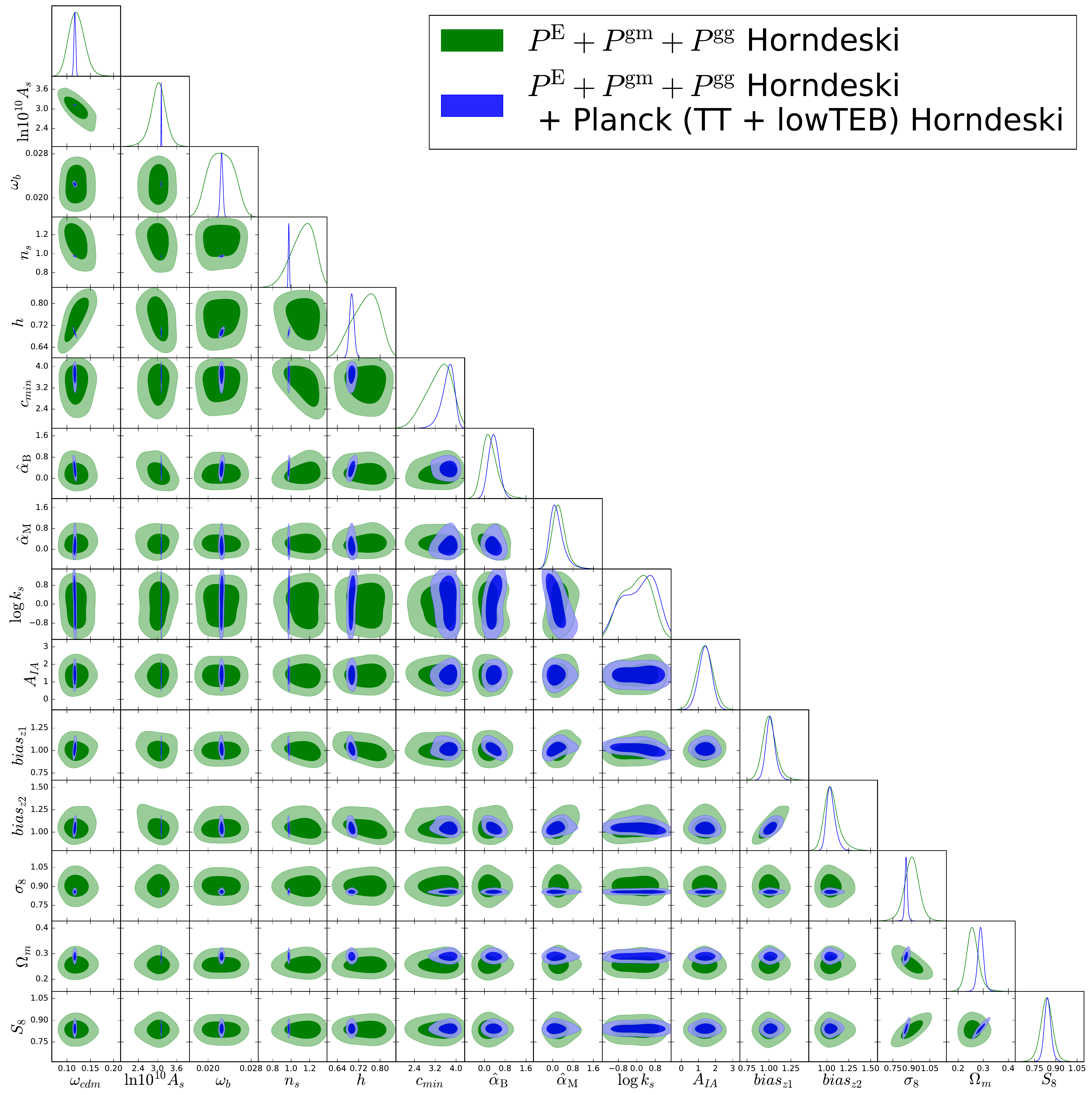}}	
	\caption{Marginalised 68 and 95 per cent contours obtained from the combination of cosmic shear, galaxy-galaxy lensing and angular clustering in Horndeski gravity (\textit{green}), and those obtained from running the same joint LSS analysis together with the \textit{Planck} CMB likelihood (\textit{blue}).}\label{fig:full_posterior}
\end{figure*}

Fig.~\ref{fig:full_posterior} shows the 68 and 95 per cent marginalised contours for all the cosmological and nuisance parameters listed in Table \ref{tab_prior}. We show the constraints that we obtain from our joint cosmic shear, galaxy-galaxy lensing and galaxy clustering analysis in Horndeski gravity with parameterization $\alpha(\tau)=\hat{\alpha} \Omega_{\mathrm{DE}}(\tau)$, as well as the contours produced from running the same LSS joint anlaysis together with \textit{Planck} likelihoods, as discussed in Section~\ref{sec:comparison_Planck}. For parameters such as $n_s$ that are not well constrained by the LSS probes considered here, the combination of the LSS likelihood with the \textit{Planck} CMB one essentially does not add much more information with respect to the one already contained in the CMB-only analysis. Conversely, for parameters such as the intrinsic alignment amplitude that are constrained only by LSS probes, the constraints do not benefit substantially from running the LSS and CMB likelihoods together. The biggest improvement that is determined by this joint LSS-CMB analysis is on parameters that are well constrained singularly by LSS and CMB datasets, such as $\Omega_{\mathrm{m}}$ and $\sigma_8$. Importantly for the analysis of this paper, this improvement also affects the Horndeski $\hat{\alpha}$ parameters.

\section{Effect of non-linearities on cosmological constraints}\label{app:nonlinearities}

\begin{figure*}
	\includegraphics[scale = 0.37]{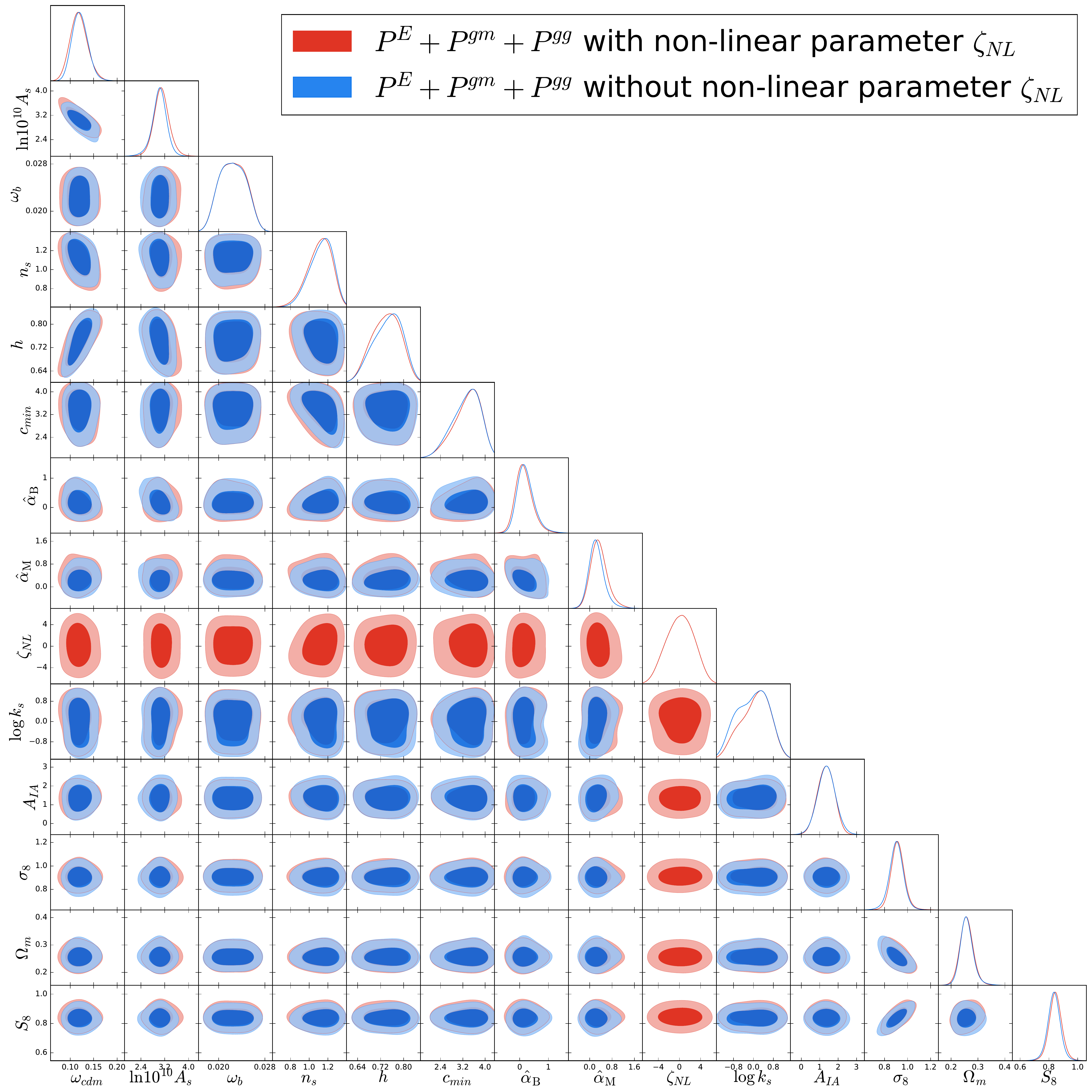}
	\caption{Marginalised 68 and 95 per cent contours obtained in our joint analysis of cosmic shear, galaxy-galaxy lensing and galaxy clustering: in \textit{red} we show the contours obtained considering the additional nuisance parameter $\zeta_{NL}$, while the \textit{blue} contours are obtained without including $\zeta_{NL}$. The contours in the two cases do not differ significantly, suggesting that our data are not deeply affected by the specific prescription used to model the non-linear power spectrum.}\label{fig:MM_allparams}
\end{figure*}

As remarked in the main text, currently there is no general prescription for the treatment of non-linear scales in cosmologies alternative to $\Lambda$CDM. Some of our data are, however, at scales affected by such non-linearities. In our analysis we decided to employ a non-linear prescription for the matter power spectrum which is the same as the one followed by the KiDS collaboration, i.e. based on the \citet{Mead2015} prescription implemented in \textsc{HMcode}, which is in turn a modification of \textsc{halofit} \citep{Smith2003}. We already warned the reader that this should be interpreted as a first order correction, meant to give an idea of the constraining power of the data, while being formally incorrect as it lacks of generality for the class of dark energy and modified gravity models considered in this analysis. To support the statement that the \citet{Mead2015} prescription is a sensitive correction to apply as a first order approach, we carried out a phenomenological study of the effect of modifications of the non-linear matter power spectrum. The idea is to parameterize our ignorance by introducing an additional nuisance parameter $\zeta_{NL}$ that quantifies the uncertainty on the non-linear matter power spectrum. This parameter only becomes `active' when the Boltzmann code \textsc{HiClass} switches to \textsc{HMcode} to source the non-linear matter power spectrum; this happens at a certain redshift-dependent scale value $k_\sigma (z)$. When this happens, a phenomenological function of scale and redshift, $\beta(k, z)$, defined as 
\begin{align}
\beta(k, z) = \log \big{[} 1 + k/k_\sigma(z) \big{]} \times \zeta_{NL},
\end{align}
modifies the non-linear matter power spectrum $P(k, z)$ sourced from \textsc{HMcode} into its `corrected' version $P_{\rm {CORR}}(k, z)$
\begin{align}
P_{\rm{CORR}}(k, z) = P(k, z) \times \big{[}1 + \beta(k, z) \big{]},
\end{align}
to account for possible deviations from $\Lambda$CDM. We vary $\zeta_{NL}$ uniformly in the range $[-5, 5]$, and marginalise over $\zeta_{NL}$ when presenting constraints on cosmological parameters. We also studied what happened when we varied the logarithm of $\zeta_{NL}$ uniformly in the range $[-5, 0]$, and found similar results. Fig.~\ref{fig:MM_allparams} shows a comparison between the cosmological constraints obtained with our joint three-probe analysis, with and without the implementation of the additional parameter $\zeta_{NL}$. We notice that the constraints on all other parameters do not change significantly, suggesting that our data are only marginally affected by different non-linear prescriptions for the matter power spectrum. Hence, our choice of using \textsc{HMcode} appears sensible and major modifications will only be required with future data at even more non-linear scales, as coming from future releases of the KiDS survey or from Stage IV surveys such as \textit{Euclid} and LSST.  

\section{Comparison with vU18}\label{app:comparison}

\begin{table*}
	\resizebox{\textwidth}{!}{%
		\begin{tabular}{|c|c|c|c|c|c|c|c|c|c|c|}
			\hline
			\hline
			Param & \multicolumn{2}{|c|}{$P^{\mathrm{E}}$} & \multicolumn{2}{|c|}{$P^{\mathrm{E}} + P^{\mathrm{gm}} + P^{\mathrm{gg}}$} & \multicolumn{2}{|c|}{$P^{\mathrm{E}} + P^{\mathrm{gm}}$} & \multicolumn{2}{|c|}{$P^{\mathrm{E}} + P^{\mathrm{gg}}$}  &  \multicolumn{2}{|c|}{$P^{\mathrm{gm}} + P^{\mathrm{gg}}$}  \\
			\hline
						 & This work & vU18 & This work & vU18 & This work & vU18 & This work & vU18 & This work & vU18 \\
			\hline
			\hline 
			$\omega_{\mathrm{cdm}}$ &  $0.139_{-0.041}^{+0.051}$ & $0.134_{-0.046}^{+0.044}$ &  $0.165_{-0.045}^{+0.033}$  & $0.153_{-0.040 }^{+0.030}$ &  $0.142_{-0.043}^{+0.032}$ & $0.132_{-0.044}^{+0.032}$ &  $0.177 _{-0.038}^{+0.035}$ & $0.169_{-0.039}^{+0.034}$ &  $0.175_{-0.051}^{+0.034}$ & $0.159_{-0.041}^{+0.035}$ \\[5pt]
			
			$\mathrm{ln}10^{10}\mathrm{A_s}$ & $2.65_{-0.92}^{+0.21}$ & $2.76_{-1.06}^{+0.30}$ &  $2.47_{-0.48}^{+0.44}$ & $2.63_{-0.44}^{+0.36}$ &  $2.63_{-0.72}^{+0.43}$ & $2.81_{-0.78}^{+0.52}$  &  $2.22_{-0.49}^{+0.14}$ & $2.30_{-0.53}^{+0.21}$ &  $2.53_{-0.48}^{+0.41}$ & $2.67_{-0.45}^{+0.37}$\\[5pt]
			
			$\omega_{\mathrm{b}}$ &    $0.0225_{-0.0033}^{+0.0030}$ & $0.0224_{-0.0034}^{+0.0036}$  &  $0.0225_{-0.0033}^{+0.0030}$  & $0.0225_{-0.0035}^{+0.0035}$ & $0.0225_{-0.0031}^{+0.0036}$ & $0.0224_{-0.0034}^{+0.0036}$ &  $0.0225_{-0.0029} ^{+0.0034}$ & $0.0225_{-0.0035}^{+0.0034}$ &  $0.0226_{-0.0029}^{+0.0035}$ & $0.0225_{-0.0033}^{+0.0034}$\\[5pt]
			
			$n_s$ &   $1.09_{-0.08}^{+0.21}$  & $1.11_{-0.05}^{+0.19}$ &  $0.95_{-0.21}^{+0.10}$ & $0.97_{-0.19}^{+0.15}$ &  $1.03_{-0.15}^{+0.20}$ & $1.08_{-0.07}^{+0.22}$ &  $0.93_{-0.19}^{+0.11}$ & $0.97_{-0.18}^{+0.14}$ &  $0.89_{-0.18}^{+0.09}$ & $0.93_{-0.22}^{+0.08}$ \\[5pt]
			
			$h$ &  $0.74_{-0.06}^{+0.07}$ & $0.74_{-0.04}^{+0.08}$ &  $0.73_{-0.07}^{+0.08}$ & $0.73_{-0.06}^{+0.09}$ &  $0.74_{-0.04}^{+0.08}$  & $0.74_{-0.03}^{+0.08}$ & $0.73_{-0.07}^{+0.07}$ & $0.73_{-0.08}^{+0.04}$  &  $0.74_{-0.05}^{+0.08}$ & $0.73_{-0.08}^{+0.09}$ \\[5pt]
			
			$c_{\mathrm{min}}$ &      $3.24_{-0.54}^{+0.74}$  & $3.27_{-0.23}^{+0.73}$ &  $2.87_{-0.79}^{+0.34}$ & $2.97_{-0.71}^{+0.56}$ &  $3.19_{-0.36}^{+0.79}$ & $3.28_{-0.22}^{+0.72}$  &  $2.98_{-0.66}^{+0.69}$ & $3.08_{-0.39}^{+0.81}$ &  $2.66_{-0.65}^{+0.25}$ & $2.86_{-0.84}^{+0.30}$\\[5pt]
			
			$A_{IA}$ &    $0.91_{-0.57}^{+0.75}$ & $0.92_{-0.59}^{+0.78}$   &  $1.24_{-0.36}^{+0.37}$ & $1.27_{-0.40}^{+0.39}$  &  $1.36_{-0.37}^{+0.36}$  & $1.46_{-0.42}^{+0.41}$  & $0.88_{-0.47}^{+0.65}$ & $0.88_{-0.50}^{+0.70}$   &  $1.38_{-0.43}^{+0.42}$ & $1.38_{-0.49}^{+0.46}$  \\[5pt]
			
			$\mathrm{bias}_{z1}$  &           -     & -            &  $1.17_{-0.18}^{+0.16}$ & $1.12_{-0.15}^{+0.14}$  &  $0.84_{-0.21}^{+0.15}$ & $0.78_{-0.19}^{+0.14}$ &  $1.25_{-0.17}^{+0.16}$ & $1.21_{-0.15}^{+0.14}$ &  $1.18_{-0.19}^{+0.18}$ & $1.13_{-0.16}^{+0.15}$ \\[5pt]
			
			$\mathrm{bias}_{z2}$ &           -   & -              &  $1.29_{-0.19}^{+0.18}$ & $1.25_{-0.17}^{+0.16}$  &  $1.55_{-0.33}^{+0.28}$  & $1.45_{-0.33}^{+0.27}$  &  $1.49_{-0.19}^{+0.20}$ &  $1.45_{-0.18}^{+0.18}$  &  $1.27_{-0.20}^{+0.19}$ & $1.23_{-0.17}^{+0.16}$\\[5pt]
			
			$\Omega_{\mathrm{m}}$ &    $0.30_{-0.08}^{+0.08}$ & $0.29_{-0.10}^{+0.07}$  &  $0.35_{-0.06}^{+0.06}$ & $0.33_{-0.06}^{+0.05}$ &  $0.30_{-0.07}^{+0.06}$ & $0.29_{-0.08}^{+0.06}$ &  $0.38_{-0.07}^{+0.06}$ & $0.36_{-0.06}^{+0.06}$ &  $0.36_{-0.08}^{+0.06}$ & $0.34_{-0.06}^{+0.05}$\\[5pt]
			
			$\sigma_8$ &     $0.77_{-0.16}^{+0.07}$  & $0.80_{-0.18}^{+0.09}$  &  $0.74_{-0.09}^{+0.06}$ & $0.78_{-0.08}^{+0.06}$ &  $0.77_{-0.12}^{+0.08}$ & $0.81_{-0.14}^{+0.09}$  &  $0.68_{-0.08}^{+0.05}$ & $0.70_{-0.08}^{+0.05}$ &  $0.77_{-0.10}^{+0.07}$ & $0.80_{-0.09}^{+0.07}$  \\[5pt]
			
			$S_8$ &         $0.760_{-0.038}^{+0.039}$  & $0.761_{-0.038}^{+0.040}$ &  $0.792_{-0.031}^{+0.032}$  & $0.800_{-0.026}^{+0.030}$ &  $0.756_{-0.035}^{+0.039}$ & $0.769_{-0.032}^{+0.037}$ &  $0.752_{-0.036}^{+0.036}$ & $0.759_{-0.032}^{+0.036}$ &  $0.840_{-0.040}^{+0.045}$ & $0.835_{-0.037}^{+0.038}$\\[5pt]
			\hline
		\end{tabular}}
		\caption{Mean and marginalised 68 per cent credibility interval on the parameters listed, obtained with our new likelihood in a $\Lambda$CDM scenario with the priors specified in Tab.~\ref{tab_prior}.}\label{tab:MyResults}
	\end{table*}

\begin{figure*}
	\includegraphics[scale=0.36]{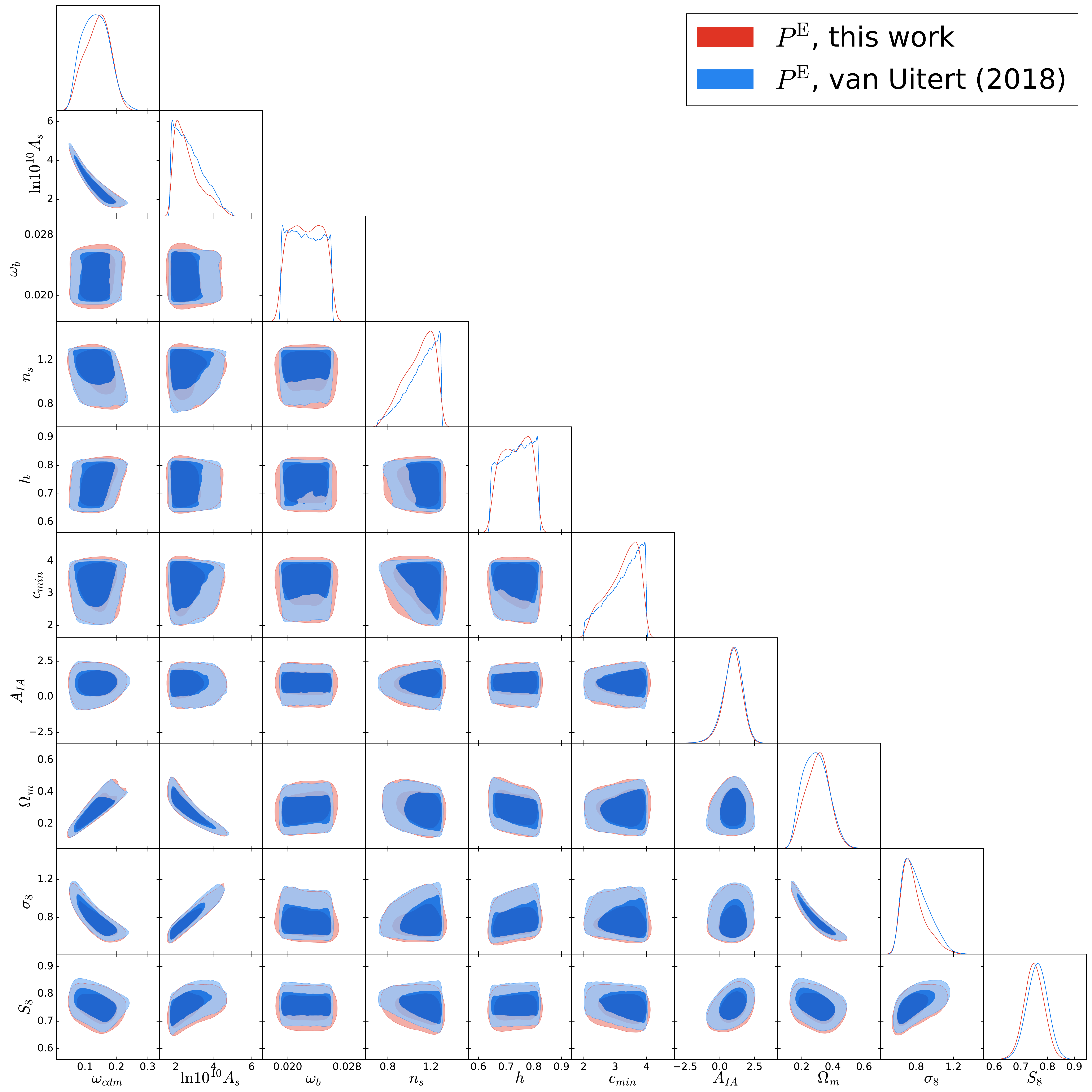}
	\caption{Comparison of the marginalised 68 and 95 per cent contours obtained in $\Lambda$CDM with our new likelihood module for the KiDS+GAMA analysis (\textit{red}) and the results by vU18 \textit{blue}. Here the probe considered is cosmic shear alone ($P^{\mathrm{E}}$).}\label{fig:comparisonMM}
\end{figure*}

\begin{figure*}
	\includegraphics[scale=0.4]{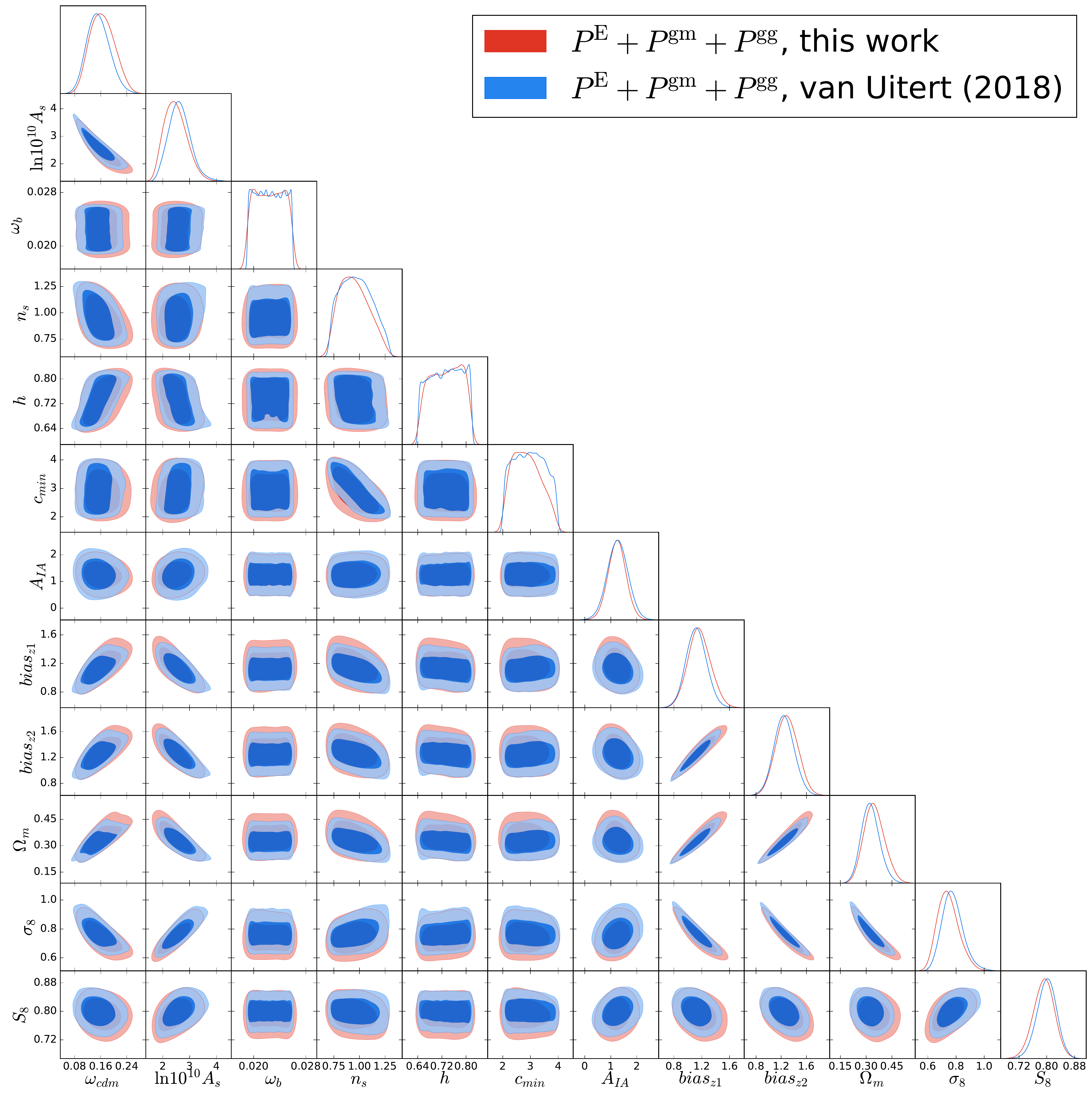}	
	\caption{Same as in Fig.~\ref{fig:comparisonMM}, but here the combination of probes considered is given by all three probes in this analysis, i.e. cosmic shear, galaxy-galaxy lensing and angular clustering ($P^{\mathrm{E}}+P^{\mathrm{GM}}+P^{\mathrm{GG}}$).}\label{fig:comparisonMMGMGG}
\end{figure*}

In Figs.~\ref{fig:comparisonMM} and \ref{fig:comparisonMMGMGG}, we select two plots of comparison with the constraints obtained by vU18, which we overplot to ours. Specifically, in the two figures we show marginalised 68 and 95 per cent contours considering cosmic shear alone and the combination of all three probes, respectively. We find excellent agreement with the results of vU18 not only with these two choices of probes combination, but for all the possible combinations. Table~\ref{tab:MyResults} summarises this comparison: it shows the mean and 68 credibility intervals obtained from our analysis and those obtained from vU18.~We notice in particular the excellent agreement in the parameter $S_8=\sigma_8\sqrt{\Omega_{\mathrm{m}}/0.3}$, derived from $\Omega_{\mathrm{m}}$ and $\sigma_8$, i.e. the two parameters whose degenerate combination cosmic shear is most sensitive to \citep{Hildebrandt2017}. Our results are obtained with a completely likelihood different implementation, which strengthens both the validity of the fiducial KiDS+GAMA analysis and the consistency of our likelihood. 

\bsp
\label{lastpage}

\end{document}